\documentclass{elsart}

\usepackage{amssymb}

\usepackage[]{latexsym}
\usepackage[]{amsfonts}
\usepackage[]{graphicx}
\usepackage{hyperref}
\usepackage{amssymb}
\usepackage{amsmath}

\newtheorem{remark1}{Remark}[section]
\newtheorem{example}{Example}
\newtheorem{definition}{Definition}

\def\boxforqed{\rule{0.5em}{1.5ex}}
\def\qed{\ifmmode\squareforqed\else{\unskip\nobreak\hfil
        \penalty50\hskip1em\null\nobreak\hfil\boxforqed
         \parfillskip=0pt\finalhyphendemerits=0\endgraf}\fi}

\def \K {{\rm I\kern -2.2pt K\hskip 1pt}}
\def \N{{\rm I\kern -2.1pt N\hskip 1pt}}
\def \R{{\rm I\kern -2.2pt R\hskip 1pt}}

\def \Q{\hskip2pt {\rm 0\kern -7pt Q}}
\def \C{{\rm C\kern -4.6pt
      \vrule height 6.8pt width 0.3pt depth -0.5pt}\hskip4pt}

\newcommand{\fraca}[2]{\displaystyle\frac{#1}{#2}}

\begin{document}

\begin{frontmatter}

%\begin{center} {\Large \bf \sc Discussing and Interpreting Linear Systems of Equalities and Inequalities. An Application to Truss Analysis}\\
%
% \vskip 0.2in
%Roberto M\'{\i}nguez$^1$, Enrique Castillo$^2$, Rosa Eva Pruneda$^1$ and Cristina Solares$^1$ \\
%
% \vskip 0.1in {\small\it $^{1}$ Department of
%Applied Mathematics, \\  University of Castilla-La Mancha, Spain }\\
% \vskip 0.1in {\small\it $^2$ Department of Applied Mathematics and Computational Sciences, \\
% University of Cantabria and University of Castilla-La Mancha, Spain}
%
%
%% \vskip 2.2in
%\end{center}



%\title{Discussing and Interpreting Linear Systems of Equalities and Inequalities.
%An Application to Truss Analysis\thanksref{label3}}
\title{Truss Analysis Discussion and Interpretation Using Linear Systems of Equalities and Inequalities}


%\author[label1,label2,label3,label4]{C. Castillo, R. M\'{\i}nguez, E. Castillo\corauthref{cor1}, M. A. Losada}

\author[label3]{R. M\'{\i}nguez\corauthref{cor1}},
\corauth[cor1]{Corresponding author. Tel.: + 34 926 810046.}
\ead[email]{rominsol@gmail.com}
%%%\ead[url]{http://www.uclm.es/profesorado/robertominguez/homepage.htm}
\author[label2]{E. Castillo},
\ead{castie@unican.es}
\author[label1]{R. Pruneda},
\ead{rosa.pruneda@uclm.es}
\author[label1]{C. Solares},
\ead{cristina.solares@uclm.es}

\address[label3]{Independent Consultant,
C/ Honduras 1, 13160 Torralba de Calatrava, Ciudad Real, Spain }

\address[label2]{Department of Applied Mathematics and Computational Sciences, Univ. de Cantabria, Spain}

\address[label1]{Department of Applied Mathematics, Univ. de Castilla-La Mancha, Spain}

\begin{abstract}
This paper shows the complementary roles of mathematical and
engineering points of view when dealing with truss analysis
problems involving systems of linear equations and inequalities.
After the compatibility condition and the mathematical structure
of the general solution of a system of linear equations is
discussed, the truss analysis problem is used to illustrate its
mathematical and engineering multiple aspects, including an
analysis of the compatibility conditions and a physical
interpretation of the general solution, and the generators of the
resulting affine space. Next, the compatibility and the
mathematical structure of the general solution of linear systems
of inequalities are analyzed and the truss analysis problem
revisited adding some inequality constraints, and discussing how
they affect the resulting general solution and many other aspects
of it. Finally, some conclusions are drawn.
\end{abstract}


\end{frontmatter}

\section{Introduction}\label{s1}There are many
engineering problems that involve linear systems of equalities and
inequalities. Engineers are familiar with
systems of equalities with a unique solution, but many of them are
not used to deal with systems of equations with multiple solutions
and their interpretations. The problem is even worse when systems
of inequalities are dealt with, because only few people know how
to obtain the associated compatibility conditions and solve them.
These two types of systems can be interpreted from the
mathematical or the engineering points of view, which are
complementary and provide a deep understanding of the problem under study.
 However, people working in these
problems use to have knowledge about only one of these two
perspectives and are unaware of the relations
between the mathematical and the engineering concepts, which leads
to important limitations in the capacity of extracting conclusions
from the results that can be expected after a careful analysis of
these problems from both points of view.

This paper points out these relations and makes them explicit for
the readers to discover the new world that arises when
contemplating the compatibility conditions or the set of general
solutions from this dual perspective.

In this paper, we have selected a particular example to illustrate these two points of view, the truss
analysis problem, and we
exploit this dual (mathematical and engineering) perspective to
deal with a problem that involves linear systems of equalities or
inequalities, depending on the constraints used to model the
reality. As we shall see, many questions of practical interest
arise and can be answered thanks to this dual analysis of the
problem.

The paper is structured as follows. In Section \ref{s3} the
problem of determining the compatibility conditions of systems of
linear equalities and solving them together with an analysis of
the general mathematical structure of their solutions is
discussed. In Section \ref{s4} the truss analysis problem is
described and used to illustrate all the theoretical methods. In
Section \ref{s5} we classify truss structures according to some
mathematical criteria in isostatic, hyperstatic, mechanism and
critical trusses. In Section \ref{s6} we discuss the compatibility
of systems of linear inequalities and the mathematical structure
of their solutions. In Section \ref{s7} we revisit the truss
analysis problem adding some constraints to illustrate the
methodology and several engineering problems. Finally, in Section
\ref{s8} some conclusions and recommendations are given.

\section{Dealing with Systems of Equations}\label{s3}In many engineering applications we
find systems of linear equations of the form:
%
%\begin{equation}\label{eq1}
%\begin{array}{ccccccc}
%a_{11}x_1 & +  a_{12}x_2 & +  \cdots  & +  a_{1n}x_n & = & b_1,\\
%a_{21}x_1 & +  a_{22}x_2 & +  \cdots  & +  a_{2n}x_n & = & b_2,\\
%\cdots  & \cdots    & \cdots  &  \cdots   &   & \cdots \\
%a_{m1}x_1 & +  a_{m2}x_2 & +  \cdots  & +  a_{mn}x_n & = & b_m,
%\end{array}
%\end{equation}
%
\begin{equation}\label{eq1}
{\bf A}{\bf x}={\bf b},
\end{equation}
where ${\bf A}\in \R^{m\times n}$, and ${\bf b}\in \R^{m}$, being $m$ and $n$ the number of equations and unknowns, respectively, which can be equal or not. Before trying to solve a system of
equations of this type it is interesting to check whether or not
the system is compatible, i. e., if it has some solutions. These
conditions can be given in terms of $b_1,b_2,\dots,b_m$, and
always have a physical or engineering meaning that raises some
light about the problem under consideration.

Once the system is proven to have solution we can obtain the set
of all its possible solutions. To solve these two problems an algorithm that gives the linear space orthogonal to the
linear space generated by a set of vectors can be used (for a
detailed description of this algorithm see Castillo, Cobo, Jubete,
Pruneda and Castillo \cite{CastilloCJPC:00}, Castillo, Cobo,
Fern\'andez-Canteli, Jubete and Pruneda \cite{CastilloCFJP:98},
Cobo, Jubete and Pruneda \cite{CastilloCJP:99}).

\subsection{Deciding whether or not a linear system of equations is compatible}\label{s3.1}
The system (\ref{eq1}) can be written as
%
%\begin{equation}\label{eq4aa}
%x_1 \left(\begin{array}{cccc}
% a_{11}\\ a_{21}\\ \vdots \\a_{m1}
%\end{array}\right)+
%x_2 \left(\begin{array}{cccc}
% a_{12}\\ a_{22}\\ \vdots \\a_{m2}
%\end{array}\right)+\cdots
%+x_n \left(\begin{array}{cccc}
% a_{1n}\\ a_{2n}\\ \vdots \\a_{mn}
%\end{array}\right) = \left(\begin{array}{cccc}
% b_1\\ b_2\\ \vdots \\b_m
%\end{array}\right),
%\end{equation}
%
\begin{equation}\label{eq4aa}
x_1 {\bf a}_1+
x_2 {\bf a}_2+\cdots
+x_n {\bf a}_n = {\bf b},
\end{equation}
which shows that the vector ${\bf b}=(b_1,\dots ,b_m)^T$ belongs
the linear space generated by the column vectors $\{{\bf a}_1,{\bf
a}_1,\dots,{\bf a}_n\}$ of the system matrix ${\bf A}$, i.e., the
compatibility requires:
$$
{\bf b}\in \mathcal {L}({\bf a}_1,{\bf a}_2,\dots,{\bf
a}_n)\Leftrightarrow {\bf b}\in \left(\mathcal{ L}({\bf a}_1,{\bf
a}_2,\dots,{\bf a}_n)^\perp\right)^\perp,
$$
where $\perp$ refers to the orthogonal set. Thus, analyzing the
compatibility of the system of equations (\ref{eq1}) reduces to
finding the linear subspace $\mathcal{ L}\left\{ {\bf w}_1,\dots ,{\bf
w}_p\right\}$ orthogonal to $\mathcal{ L}\left\{ {\bf a}_1,\dots ,{\bf
a}_n\right\}$ and checking whether or not ${\bf b}^T{\bf W}={\bf
0}$.

\begin{example}[Compatibility of a linear system of equations]\label{ex4.10}{\em Suppose that we are interested in determining the conditions under
which the system of equations

\begin{equation}\label{comp1}\begin{array}{rrrrrcr}
x_1 & + x_2 &- x_3 & + x_4 & + x_5& = &a \\
& -x_2 & + 2 x_3 & + x_4 &- 2 x_5& = &b\\
 x_1 &- x_2 & + 2 x_3& & & = &c\\
 -2 x_1 & + 2 x_2 &- 3 x_3 & + x_4 &- x_5& = &d\\
 x_1 & + x_2 & + x_3 & + x_4 & + x_5& = &e\\
\end{array}
\end{equation}
is compatible. Then, first we obtain the linear subspace
orthogonal to the linear subspace generated by the column vectors
in (\ref{comp1}) that is:
\begin{equation}
{\bf W}= \mathcal{ L}\left\{{\bf w}_1\right\}=\mathcal{ L}\left\{(0,2,-7,-3,1)^T\right\},
\end{equation}
which implies the following compatibility condition:
\begin{equation}
\begin{array}{lcl}
{\bf w}_1 ^T(a,b,c,d,e)^T & = & (0,2,-7,-3,1) (a,b,c,d,e)^T=0
\Rightarrow 2b-7 c-3d +e=0.
\end{array}\end{equation}
}
\end{example}

\subsection{Solving a homogeneous system of linear equations}\label{s5.6}Consider the homogeneous system of
equations
%
%\begin{equation}\label{eq1a}
%\begin{array}{cccccc}
%a_{11}x_1 & +  a_{12}x_2 & +  \cdots  & +  a_{1n}x_n & = & 0,\\
%a_{21}x_1 & +  a_{22}x_2 & +  \cdots  & +  a_{2n}x_n & = & 0,\\
%\cdots\cdots  & \cdots\cdots    & \cdots\cdots   &  \cdots\cdots   &   & \cdots \\
%a_{m1}x_1 & +  a_{m2}x_2 & +  \cdots  & +  a_{mn}x_n & = & 0\\
%\end{array}
%\end{equation}
%
\begin{equation}\label{eq1a}
{\bf A}{\bf x}={\bf 0},
\end{equation}
which can be written as
%
%\begin{equation}\label{eq5aa}
%\begin{array}{ccc}
% (a_{11}, \ldots ,a_{1n})  (x_1, \ldots , x_n)^T & = & 0,\\
% (a_{21}, \ldots ,a_{2n})  (x_1, \ldots , x_n)^T & = &0,\\
%  \cdots \cdots \cdots \cdots \cdots\cdots\cdots\cdots \cdots   &  & \cdots\\
% (a_{m1}, \ldots ,a_{mn})  (x_1, \ldots , x_n)^T & = & 0.\\
%\end{array}
%\end{equation}
%
\begin{equation}\label{eq5aa}
{\bf a}^i {\bf x}^T = 0;\;i=1,\ldots,m.
\end{equation}

Expression (\ref{eq5aa}) shows that $( x_1,\dots ,x_n)$ is
orthogonal to the set of row vectors $\{{\bf a}^1,{\bf
a}^2,\dots,{\bf a}^m\}$ of ${\bf A}$.

Then, obtaining the solution to system (\ref{eq5aa}) reduces to
determining the linear subspace orthogonal to the linear subspace
generated by the rows of matrix ${\bf A}$. Thus, the general
solution of an homogeneous system of linear equations is a linear
space, i.e., of the form
%
%$$
%\left(\matrix{x_1\cr x_2\cr \vdots\cr  x_n\cr
%}\right)=\sum\limits_{i=1}^p \rho_i\left(\matrix{v_{i1}\cr
%v_{i2}\cr \vdots\cr v_{in}\cr }\right);\quad \rho_i\in \R.
%$$
$$
{\bf x}=\sum\limits_{i=1}^p \rho_i{\bf v}_i;\quad \rho_i\in \R.
$$

\begin{example}[An homogeneous system of linear equations] \label{ex11}{\em
The system of linear equations
\begin{equation}\label{ex1aa}\begin{array}{rrrrrrl}
x_1 & & + x_3 & - 2 x_4 & + x_5  &=&  0\\
x_1 & - x_2 & - x_3 & + 2 x_4 & + x_5 &=&  0\\
-x_1 & + 2 x_2 & + 2 x_3 & -  3 x_4 & + x_5&=&  0\\
\end{array}
\end{equation}
has as general solution the linear space
$$\left(\begin{matrix}x_1\cr x_2\cr x_3\cr x_4\cr x_5\cr \end{matrix}\right)=\rho_1\left(\begin{matrix}1\cr 2\cr -3\cr -1\cr 0\cr\end{matrix}\right)+\rho_2
\left(\begin{matrix} 0 \cr  2 \cr -7\cr -3\cr  1\cr \end{matrix}\right);\quad
\rho_1,\rho_2\in \R.
 $$
}
\end{example}

\subsection{Solving a complete system of linear equations}\label{s5.7}
Now consider again the complete system of linear equations
(\ref{eq1}), that adding the artificial variable $x_{n+1}$, can be
written as
\begin{equation}\label{eq2aaa}
\begin{array}{ccccrrc}
a_{11}x_1 & +  a_{12}x_2 & +  \cdots  & +  a_{1n}x_n & - b_1x_{n+1} & = & 0,\\
a_{21}x_1 & +  a_{22}x_2 & +  \cdots  & +  a_{2n}x_n & - b_2x_{n+1}& = & 0,\\
\cdots  & \cdots    & \cdots   &  \cdots   & \cdots    &   & \cdots  \\
a_{m1}x_1 & +  a_{m2}x_2 & +  \cdots  & +  a_{mn}x_n & - b_mx_{n+1}& = & 0,\\
 & & & &x_{n+1}& = & 1.
\end{array}
\end{equation}

The first $m$ equations of the system (\ref{eq2aaa}) can be
written as
\begin{equation}\label{eq5aab}
\begin{array}{ccc}
 (a_{11}, \ldots ,a_{1n},-b_1)  (x_1, \ldots , x_n, x_{n+1})^T & = & 0,\\
 (a_{21}, \ldots ,a_{2n},-b_2)  (x_1, \ldots , x_n, x_{n+1})^T & = &0,\\
  \cdots \cdots \cdots \cdots \cdots\cdots\cdots\cdots \cdots \cdots \cdots \ &  & \cdots\\
 (a_{m1}, \ldots ,a_{mn},-b_m)  (x_1, \ldots , x_n, x_{n+1})^T & = & 0,\\
\end{array}
\end{equation}
which shows that $( x_1,\dots ,x_n, x_{n+1})$ is orthogonal to the
set of vectors
$$\{( a_{11},\dots , a_{1n},-b_1),( a_{21},\dots , a_{2n},-b_2),
\dots , ( a_{m1},\dots , a_{mn},-b_m)\},$$
i.e., the solution of
(\ref{eq5aab}) is the linear subspace orthogonal to the linear
subspace generated by the rows of matrix ${\bf A}_b$, i.e.,
$$
\mathcal{ L}\{ (a_{11},\dots , a_{1n},-b_1),(a_{21},\dots ,
a_{2n},-b_2), \dots , (a_{m1},\dots  ,a_{mn},-b_m) \}^{\perp}.
$$

Thus, the solution of (\ref{eq1}) is the projection on $X_1\times
\cdots \times X_n$ of the intersection of such a orthogonal linear
subspace and the set $\{{\bf x}|x_{n+1}=1\}$. In other words, the
general solution of a complete system of linear equations is an
affine space, that is the sum of a constant vector plus a linear
space:
%
%$$
%\left(\matrix{x_1\cr x_2\cr \vdots\cr  x_n\cr
%}\right)=\left(\matrix{x_1^0\cr x_2^0\cr \vdots\cr  x_n^0\cr
%}\right)+\sum\limits_{i=1}^p \rho_i\left(\matrix{v_{i1}\cr
%v_{i2}\cr \vdots\cr v_{in}\cr }\right);\quad \rho_i\in \R,
%$$
%
$$
{\bf x}={\bf x}^0+\sum\limits_{i=1}^p \rho_i{\bf v}_i;\quad \rho_i\in \R,
$$
where the first vector in the right hand side is an arbitrary
particular solution of the system (\ref{eq1}), and the linear
space is the set of solutions of the associated homogeneous
system.

\begin{example}[A complete system of linear equations] {\em
The system of linear equations
\begin{equation}\label{ex1aaz}\begin{array}{rrrrrrl}
x_1 & & + x_3 & - 2 x_4 & + x_5  &=&  1\\
x_1 & - x_2 & - x_3 & + 2 x_4 & + x_5 &=&  2\\
-x_1 & + 2 x_2 & + 2 x_3 & -  3 x_4 & + x_5&=&  1\\
\end{array}
\end{equation}
has as general solution the affine space
$$\left(\begin{matrix}x_1\cr x_2\cr x_3\cr x_4\cr x_5\cr \end{matrix}\right)=\left(\begin{matrix}1\cr 1\cr 1\cr 1\cr 1\cr \end{matrix}\right)+\rho_1\left(\begin{matrix}1\cr 2\cr -3\cr -1\cr 0\cr\end{matrix}\right)+\rho_2
\left(\begin{matrix}  0 \cr  2 \cr -7\cr -3\cr  1\cr  \end{matrix}\right);\quad
\rho_1,\rho_2\in \R.
 $$
}
\end{example}

\begin{figure}[h]
\begin{center}
\includegraphics*[scale=0.50]{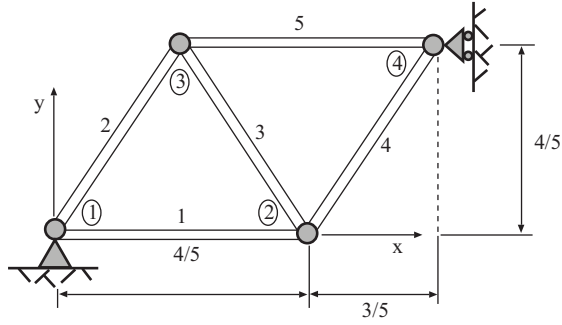}
\caption{Two-dimensional truss structure consisting of bars joined by frictionless hinges.}\label{f1}
\end{center}
\end{figure}

\section{The Truss Analysis Problem}\label{s4}
Consider the engineering problem of structures consisting of
trusses, i.e., structures made of bars joined by frictionless
hinges which imply that only tension or compression stresses exist in the
bars. A typical truss structure is shown in Figure \ref{f1}.

\begin{figure}[h]
\begin{center}
\includegraphics*[scale=0.50]{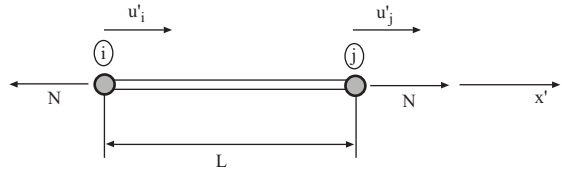}
\caption{Loads acting on an individual bar.}\label{f4}
\end{center}
\end{figure}

With reference to the bar element (see Figure \ref{f4}) that has
length $L$ and cross-section $A$, let us first establish the
stiffness $k$. By definition, the force $N$ is related to the
normal stress $\sigma$ as
\begin{equation}\label{equ1}
N=A\sigma,
\end{equation}
and assuming that the bar behavior is linearly elastic, according
to Hooke's law, we have
\begin{equation}\label{equ2}
\sigma=E\epsilon,
\end{equation}
where $E$ is the Young's modulus and $\epsilon$ is the
longitudinal strain, that is assumed constant along its length,
and has value
\begin{equation}\label{equ3}
\epsilon=\fraca{u'_j-u'_i}{L},
\end{equation}
where $u'_i$ and $u'_j$ are the longitudinal displacements of the
end points $i$ and $j$, respectively.

Combining Equations (\ref{equ1})-(\ref{equ3}) we obtain
\begin{equation}\label{equ4}
N=k(u'_j-u'_i),
\end{equation}
where $k=AE/L$ is the bar stiffness.

\begin{figure}[h]
\begin{center}
\includegraphics*[scale=0.50]{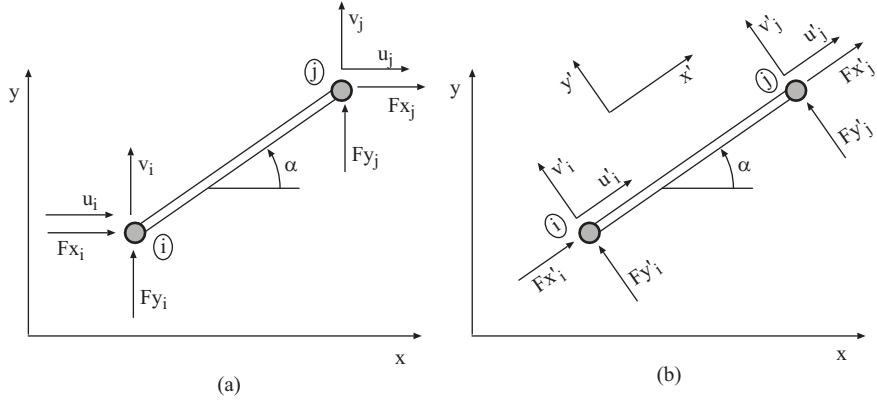}
\caption{Displacements and force components of a bar element with
respect to: (a) the global $x$-$y$ system, and (b) the local
$x'$-$y'$ system.}\label{f5}
\end{center}
\end{figure}

To calculate a given truss structure first we need to know the
behavior of a single bar in an arbitrary direction. Thus, consider
the bar shown in Figure \ref{f5} (a). Our purpose is to determine
the relation between the node forces
$F_{x_i},F_{y_i},F_{x_j},F_{y_j}$ and the node displacements $u_i,
v_i, u_j,v_j$.

To this aim, we introduce a local $x'$-$y'$-coordinate system (see
Figure \ref{f5} (b)) with its origin at the end point $i$ and its
axis $x'$ directed along the bar axis from the end point $i$
towards the end point $j$. Note that this coordinate system can be
obtained from the $xy$-system by a rotation of angle $\alpha$.

From Figures \ref{f4} and \ref{f5} (b) and Equation (\ref{equ4})
we obtain
\begin{equation}\label{equ5}
\begin{array}{rcl}
  F'_{x_i}=-N & = & k (u'_i-u'_j) \\
  F'_{x_j}=N & = & k (u'_j-u'_i),
\end{array}
\end{equation}
which can be written as
\begin{equation}\label{equ6}
{\bf F}^{e'}=\left(
\begin{array}{c}
  F'_{x_i} \\
  F'_{y_i} \\
  F'_{x_j} \\
  F'_{y_j} \\
\end{array}\right)={\bf K}^{e'}{\bf u}^{e'}=k
\left(
\begin{array}{cccc}
  1 & 0 & -1 & 0 \\
  0 & 0 &  0 & 0 \\
  -1 & 0 & 1 & 0 \\
  0 & 0 &  0 & 0 \\
\end{array}\right)
\left(
\begin{array}{c}
  u'_i \\
  v'_i \\
  u'_j \\
  v'_j \\
\end{array}\right),
\end{equation}
where ${\bf K}^{e'}$ is the so-called local element stiffness
matrix, that is symmetric. It can be observed that the
displacements along the $y'$-axis, i.e., $v'_i$ and $v'_j$,
produce no forces on the bar. This assumption is valid only if the
displacements are small; otherwise, the components $v'_i$ and
$v'_j$ might result in an elongation of the bar and thus in the
development of forces.

Let us now establish a relation between the displacements $u_i,
v_i, u_j,v_j$ and $u'_i, v'_i, u'_j,v'_j$. By geometrical
arguments it follows directly that
\begin{equation}\label{equ7}
\begin{array}{ccc}
  u_i & = & u'_i \cos \alpha-v'_i \sin \alpha \\
  v_i & = & u'_i \sin \alpha+v'_i \cos \alpha
\end{array}
\end{equation}
and the corresponding equations for $u_j$ and $v_j$.

 Therefore
\begin{equation}\label{equ7a}
{\bf u}^{e}=\left(
\begin{array}{c}
  u_i \\
  v_i \\
  u_j \\
  v_j \\
\end{array}\right)={\bf L}^{e^T}{\bf u}^{e'}=
\left(
\begin{array}{cccc}
  \cos\alpha & -\sin\alpha & 0 & 0 \\
  \sin\alpha & \cos\alpha &  0 & 0 \\
  0 & 0 & \cos\alpha & -\sin\alpha \\
  0 & 0 &  \sin\alpha & \cos\alpha \\
\end{array}\right)
\left(
\begin{array}{c}
  u'_i \\
  v'_i \\
  u'_j \\
  v'_j \\
\end{array}\right),
\end{equation}
where ${\bf L}^{e}$ is the so-called transformation matrix, that
is an orthogonal matrix (${\bf L}^{e^{-1}}={\bf L}^{e^T}$).

Premultiplying (\ref{equ7a}) by ${\bf L}^{e}$ and using the
orthogonal property, we then conclude that
\begin{equation}\label{equ8}
{\bf u}^{e'}={\bf L}^{e}{\bf u}^{e}.
\end{equation}

As both forces and displacements are vector quantities, we can
choose the same basis for the forces and then, the element forces
${\bf F}^{e}$ and ${\bf F}^{e'}$ are related in exactly the same
manner as the element displacements, i.e.,
\begin{equation}\label{equ9}
{\bf F}^{e}={\bf L}^{e^T}{\bf F'}^{e}.
\end{equation}

With these preliminary remarks, we can determine the relation
between the element forces ${\bf F}^{e}$ and the element
displacements ${\bf u}^{e}$. Insertion of (\ref{equ8}) into
(\ref{equ6}) yields
\begin{equation}\label{equ10}
{\bf F}^{e'}={\bf K}^{e'}{\bf L}^{e}{\bf u}^{e},
\end{equation}
and premultiplication by ${\bf L}^{e^T}$ and using (\ref{equ9})
leads to
\begin{equation}\label{equ11}
{\bf F}^{e}={\bf L}^{e^T}{\bf F'}^{e}={\bf L}^{e^T}{\bf
K}^{e'}{\bf L}^{e}{\bf u}^{e}={\bf K}^{e}{\bf u}^{e},
\end{equation}
where ${\bf K}^{e}$ is the so-called global element stiffness
matrix. Its explicit expression is
\begin{equation}\label{equ12}
{\bf K}^{e}=k \left(
\begin{array}{cccc}
  \cos^2\alpha & \cos\alpha\sin\alpha & -\cos^2\alpha & -\cos\alpha\sin\alpha \\
  \cos\alpha\sin\alpha & \sin^2\alpha &  -\cos\alpha\sin\alpha & -\sin^2\alpha \\
  -\cos^2\alpha & -\cos\alpha\sin\alpha & \cos^2\alpha & \cos\alpha\sin\alpha\\
  -\cos\alpha\sin\alpha & -\sin^2\alpha &  \cos\alpha\sin\alpha & \sin^2\alpha\\
\end{array}\right),
\end{equation}
which shows the symmetric character of the global element stiffness
matrix ${\bf K}^{e}$.

The truss analysis problem has the following elements to be
considered:

\begin{description}
\item[{\bf Bars: }] The longitudinal elements supporting only
tension or compression stresses (the number of bars is $b$).

\item[{\bf Nodes: }] The frictionless elements that join the bars
allowing relative rotation between them (the number of hinges or
nodes is $m$).

\item[{\bf Supports: }] Elements that prevent the structure to
experiment solid rigid displacements. The reactions exerted by the
supports are the necessary ones for the equilibrium to hold. The
number of constraints associated with the supports is denoted $c$.

\item[{\bf Forces: }] The forces acting on the nodes (hinges) of
the structure that are the data of our problem.

\item[{\bf Displacements: }] Movements suffered by the nodes owing
to the strains produced by the stresses in the bars.

  \item[{\bf Unknowns: }]  The hinge (nodal) displacements. The number of unknowns
coincides with the number of possible hinge displacements, which is
the number of degrees of freedom (two times the number of nodes
$m$ in case of bi-dimensional analysis).

  \item[{\bf Equations: }] The mathematical relations that give the nodal forces
  in terms of the nodal displacements. To derive the system of equations that model
  a given problem the global structure stiffness matrix must be obtained.
\end{description}

\begin{figure}[h]
\begin{center}
\includegraphics*[scale=0.50]{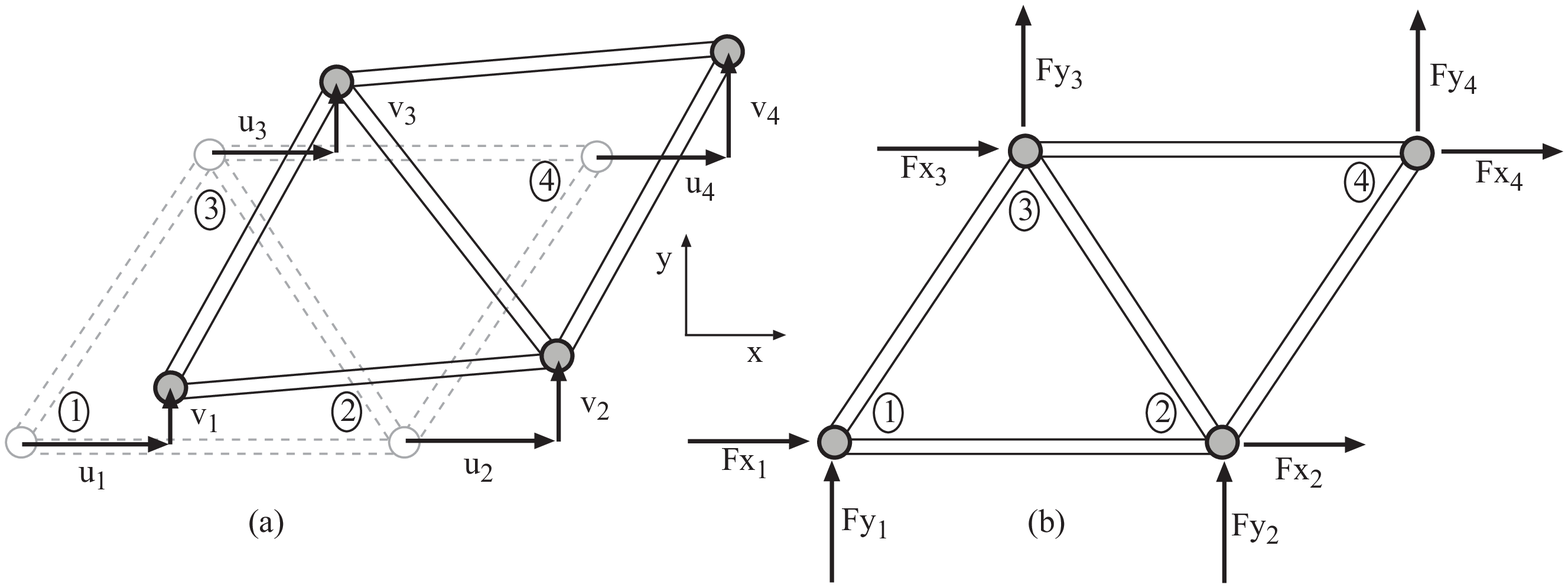}
\caption{Displacements and force components for the
two-dimensional truss structure.}\label{f2}
\end{center}
\end{figure}

\begin{example}[A simple truss structure]{\em As an example of how the element stiffness matrix can be used to
derive the stiffness matrix for the whole structure, consider the
simple truss structure shown in Figure \ref{f1}. Using the
relationships between forces and movements, establishing the
equilibrium of vertical and horizontal forces for each node, and
replacing the static magnitudes (stresses) by their equivalents as
a function of the nodal displacements, then
the so-called global assembled stiffness matrix is obtained. Note
that each element stiffness matrix (${\bf K}^{e}$) is assembled
into the global element stiffness matrix (${\bf K}$), to get the
following system of equations:
\begin{equation}\label{e100}
{\bf K}{\bf u}\!=\!\fraca{k}{150}\!\!
\left(\renewcommand{\arraycolsep}{0.00cm}\renewcommand{\arraystretch}{1.1}\begin{array}{ccccccccccc}
179 & 72 & | &  -125 & 0 & | &  -54 & -72 & | &  0 & 0\\
72 & 96 & | &  0 & 0 & | &  -72 & -96 & | &  0 & 0\\
- & - & + & - & - & + & - & - & + & - & - \\
-125 & 0 & | &  233 & 0 & | &  -54 & 72 & | &  -54 & -72\\
0 & 0 & | &  0 & 192 & | &  72 & -96 & | &  -72 & -96\\
- & - & + & - & - & + & - & - & + & - & - \\
-54 & -72 & | &  -54 & 72 & | &  233 & 0 & | &  -125 & 0\\
-72 & -96 & | &  72 & -96 & | &  0 & 192 & | &  0 & 0\\
- & - & + & - & - & + & - & - & + & - & - \\
0 & 0 & | &  -54 & -72 & | &  -125 & 0 & | &  179 & 72\\
0 & 0 & | &  -72 & -96 & | &  0 & 0 & | &  72 & 96
\end{array}\right)\!\!\!\left(\begin{array}{c}
 u_{1} \cr  v_{1} \cr  -- \cr u_{2} \cr  v_{2} \cr -- \cr  u_{3} \cr  v_{3}  \cr -- \cr u_{4} \cr  v_{4} \cr
\end{array}\right)\!=\!\left(\begin{array}{c}
F_{x_1} \cr F_{y_1} \cr -- \cr F_{x_2}  \cr F_{y_2}  \cr -- \cr
F_{x_3} \cr F_{y_3} \cr -- \cr F_{x_4}  \cr F_{y_4}  \cr
\end{array}\right)\!\!={\bf F}\renewcommand{\arraystretch}{1}
\end{equation}
where $u_i, v_i; i=1,\ldots,4$ are the node displacements shown in
Figure \ref{f2} (a) and $F_{x_i}, F_{y_i}; i=1,\ldots,4$ are the
external forces acting on the hinges of the structure including
support reactions (see Figure \ref{f2} (b)). The different
partitions refer to different nodes.

First, using the method described
  in Section \ref{s3.1} the conditions to be satisfied
 by the linear system (\ref{e100}) to have solution (one or more) are obtained:
 \begin{eqnarray}
 0 &=& F_{x_1}+F_{x_2}+F_{x_3}+F_{x_4} \label{equ13}\\
 0 &=&   F_{y_1}+F_{y_2}+F_{y_3}+F_{y_4} \label{equ14}\\
 0 &=&   4/5 F_{x_1}-3/5 F_{y_1}+4/5 F_{x_2}+3/5 F_{y_2}+6/5 F_{y_4},\label{equ15}
 \end{eqnarray}
where (\ref{equ13}) and (\ref{equ14}) express the equilibrium of
horizontal and vertical forces respectively, and (\ref{equ15})
establishes the equilibrium of moments (the moment are taken with
respect to node $3$). Thus, these conditions state that the
structure subject to the external forces and the support reactions
must be in equilibrium ($\sum F_{x_i}=0, \sum F_{y_i}=0$ and $\sum
M_i=0$).

The rank of ${\bf K}$ in (\ref{e100}) is $5$ that implies that
three ($8-5$) rows are linear combinations of the other rows, and then the
system (\ref{e100}) has infinite solutions. So, some extra
equations of the form ${\bf B}{\bf u}={\bf b}$,
%
%$
%{\bf B}\left(\begin{array}{c}
% u_{1} \cr  v_{1} \cr  u_{2} \cr  v_{2} \cr  u_{3} \cr  v_{3}  \cr u_{4} \cr  v_{4} \cr
%\end{array}\right)={\bf b},
%$
%
where ${\bf b}$ is the boundary conditions vector, can be added
maintaining the compatibility of system (\ref{e100}).

To obtain the set of all possible solutions of system
(\ref{e100}), we apply the orthogonalization algorithm and obtain:
\begin{equation}\label{e105}
\left(\begin{array}{c}
 u_{1} \cr  v_{1} \cr  u_{2} \cr  v_{2} \cr  u_{3} \cr  v_{3} \cr  u_{4} \cr  v_{4} \cr
\end{array}\right)=\fraca{1}{k}\left(\begin{array}{c}
 467/120 \cr  501/160 \cr  359/120 \cr  -359/160 \cr  39/10 \cr  0 \cr  0 \cr  0 \cr
\end{array}\right)+\rho_1\left(\begin{array}{c}
1 \cr  0 \cr  1 \cr  0 \cr  1 \cr  0 \cr  1 \cr 0 \cr
\end{array}\right)+\rho_2\left(\begin{array}{c}
0 \cr  1 \cr  0 \cr  1 \cr  0 \cr  1 \cr  0 \cr 1 \cr
\end{array}\right)+\rho_3\left(\begin{array}{c}
4/5 \cr  -3/5 \cr  4/5 \cr  3/5 \cr  0 \cr  0 \cr  0 \cr 6/5 \cr
\end{array}\right);\quad \rho_1,\rho_2,\rho_3\in\R,
\end{equation}
which from a mathematical point of view is an affine space, i.e.,
the sum of a given vector (the first one) plus a linear space of
dimension $3$ (an arbitrary linear combination of 3 linearly
independent vectors).

\begin{figure}[h]
\begin{center}
\includegraphics*[scale=0.6]{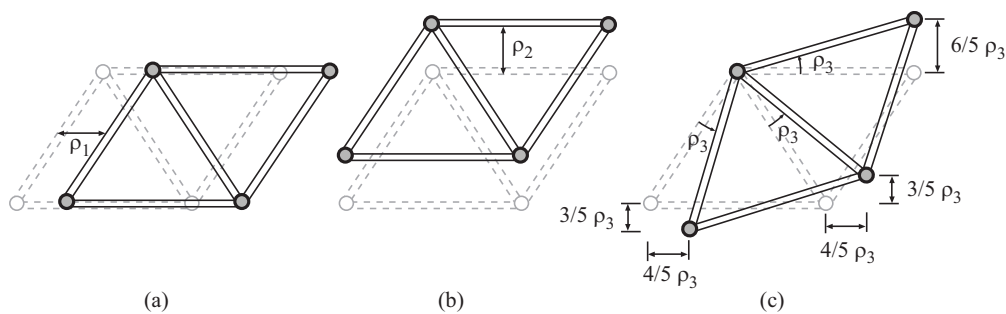}
\caption{Illustration of the basis of the linear space of
dimension three that appears in the general solution of the
two-dimensional truss structure in Example \ref{ex32}. The basis
has three generators that correspond to: (a) a horizontal
translation, (b) a vertical translation, and (c) a
rotation.}\label{f3}
\end{center}
\end{figure}

Note that the linear space is generated by three vectors that, due
to the symmetry of the stiffness matrix, coincide with those in
(\ref{equ13})-(\ref{equ15}). From an engineering point of view,
this solution must be interpreted as follows:
\begin{enumerate}

\item The dimension of the linear space is the number of degrees
of freedom we have, i.e., the maximum number of $u$ and $v$
displacements that can be fixed for the system to have a unique
solution.

\item The linear space component of the solution (unlimited values
of the $\rho$ coefficients, and consequently unlimited
displacements) implies that the structure can be located anywhere
whereas the forces acting on the structure guarantee equilibrium.

  \item The first vector is a particular solution, i.e., a
  solution to the stated problem. Note that it satisfies Equation
  (\ref{e100}) for the ${\bf F}$ values in Figure \ref{f6}. It is worth noting that this
  vector can be replaced by any other particular solution, which can be obtained by adding to it any
linear combination of the three basic vectors.

  \item The second vector corresponds to a solution of the
  associated homogeneous problem, i.e., with no external forces. In this particular case it corresponds
  to the horizontal displacement of
  the structure as a rigid solid (see Figure \ref{f3}(a)).

  \item The third vector corresponds to another solution of the
  associated homogeneous problem. In this case, it is the vertical displacement of
  the structure as a rigid solid (see Figure \ref{f3}(b)).

\item Finally, the fourth vector corresponds to another solution
of the associated homogeneous problem, which corresponds to the
rotation of the structure as a rigid solid with respect node $3$
(see Figure \ref{f3}(c)).
\end{enumerate}

It is obvious that the linear space generated by the last three
vectors in (\ref{e105}) can be represented using another basis of
the same space, i.e., considering equilibrium with respect other point.
}
\end{example}

\section{Classification of truss structures}\label{s5}
Structures can be classified in different ways depending on the
rank of the matrix $\left(\begin{array}{ccc}
  {\bf K} & | &
  {\bf B} \\
\end{array}\right)^T$ and the value of $b+c$ and $m$, as shown in Table
\ref{t900}. They include isostatic, hyperstatic, critical and
mechanism truss structures:
\begin{description}
    \item[Isostatic:] In this kind of structures $2m=b+c$ and the $\mbox{rank}\left(\begin{array}{ccc}
  {\bf K} & | &
  {\bf B} \\
\end{array}\right)^T=2m$, i.e., a unique solution exists. They are characterized because (a) the boundary conditions ensure equilibrium under all possible
external forces, (b) thermic strains do not induce stresses, and (c) if any of their elements (nodes, bars, or boundary conditions) is removed the structure becomes a mechanism or a critical structure.

    \item[Hyperstatic:] In this kind of structures $b+c>2m$ where $(b+c)-2m$ is the degree of hyperstaticity and the $\mbox{rank}\left(\begin{array}{ccc}
  {\bf K} & | &
  {\bf B} \\
\end{array}\right)^T=2m$, i.e., a unique solution exists. They are characterized because (a) the boundary conditions ensure equilibrium under all possible
external forces, (b) thermic strains induce stresses, and (c) if any of their elements (nodes, bars, or boundary conditions) is removed the structure becomes a isostatic structure or remains hyperstatic. Note that there are two hyperstatic
cases, one is due to an excess of bars in the truss structure, and
the other is due to an excess in the supports (boundary
conditions).
    \item[Mechanism:] In this kind of structures $b+c\le 2m$ and the $\mbox{rank}\left(\begin{array}{ccc}
  {\bf K} & | &
  {\bf B} \\
\end{array}\right)^T<2m$, i.e., infinite solution exists. They are characterized because (a) the boundary conditions do not ensure equilibrium under all possible
external forces, (b) thermic strains do not induce stresses.
    \item[Critical:] In this kind of structures $2m\ge b+c$ and the $\mbox{rank}\left(\begin{array}{ccc}
  {\bf K} & | &
  {\bf B} \\
\end{array}\right)^T<2m$, i.e., infinite solution exists. They are characterized because (a) the boundary conditions do not ensure equilibrium under all possible
external forces within the hypothesis of small deformations, (b) thermic strains induce stresses, and (c) if any of their elements (nodes, bars, or boundary conditions) is removed the structure becomes a mechanism.
   \end{description}

To illustrate their main differences we include some examples.

\begin{table}
  \centering
  \caption{A classification of truss structures.}\label{t900}
  \vspace*{0.3cm}
  \begin{tabular}{|c|c|c|}
      \hline
     & Rank$\left(\begin{array}{c}
  {\bf K} \\
 - \\
  {\bf B} \\
\end{array}\right)$ & $b+c$\\
    \hline
    Isostatic & $2m$ & $2m$\\
        \hline
    Hyperstatic & $2m$ & $>2m$ \\
        \hline
    Critical & $<2m$ & $\ge 2m$ \\
        \hline
    Mechanism & $<2m$ & $\le 2m$ \\
    \hline
  \end{tabular}
\end{table}

%\begin{table}
%  \centering
%  \caption{A classification of truss structures.}
%  \vspace*{0.3cm}
%  \begin{tabular}{|c|c|c|c|}
%      \hline
%    % after \\: \hline or \cline{col1-col2} \cline{col3-col4} ...
%     & Rank$\left(\begin{array}{c}
%  {\bf K} \\
% - \\
%  {\bf B} \\
%\end{array}\right)$ & $b+c$ & Compatibility conditions\\
%    \hline
%    Isostatic & $2m$ & $2m$& $3$\\
%        \hline
%    Hyperstatic (boundary conditions)& $2m$ & $>2m$ & $>3$ \\
%        \hline
%    Hyperstatic (bars)& $2m$ & $>2m$ & $3$ \\
%        \hline
%    Critical & $<2m$ & $>2m$ & $3$ \\
%        \hline
%    Mechanism & $<2m$ & $\le 2m$ & $<3$\\
%    \hline
%  \end{tabular}
%\end{table}

\begin{example}[An isostatic truss structure]\label{ex32}{\em
Consider the particular case shown in Figure \ref{f6}, where there
are $3$ external forces. Since the vector of external forces ${\bf
F}$ has to fulfill the compatibility conditions
(\ref{equ13})-(\ref{equ15}), we use them to obtain the truss
reactions:

\begin{figure}[h]
\begin{center}
\includegraphics*[scale=0.50]{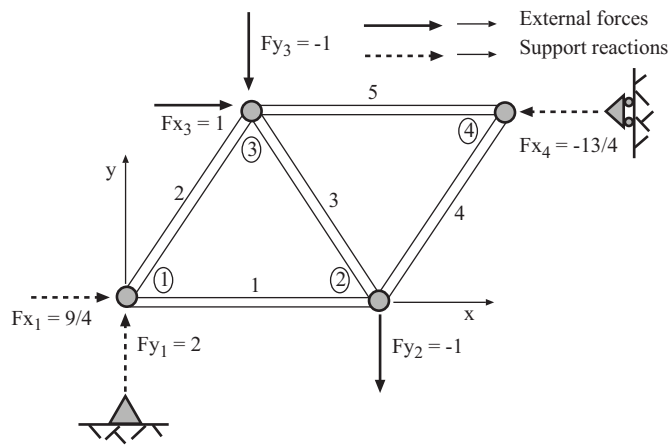}
\caption{The two-dimensional truss structure showing the external
forces and reactions.}\label{f6}
\end{center}
\end{figure}

$$
F_{x_1}=9/4;\quad F_{y_1}=2;\quad F_{x_4}=-13/4;\quad
$$
and then, the force vector becomes: ${\bf F}=1/4\left(\begin{matrix}9 & 8 & 0 & -4 & 4 &
-4 & -13 & 0 \cr \end{matrix}\right)^T.$

Next, we consider the boundary conditions imposed by the supports
that establish the final location of the structure. Assuming that
those boundary conditions are $u_1=0, v_1=0$ and $u_4=0$, then
from (\ref{e105}) we have:
\begin{eqnarray*}
    u_1 & = & 467/(120k)+\rho_1+4/5\rho_3 = 0  \\
    v_1 & = & 501/(160k)+\rho_2-3/5\rho_3 = 0  \\
    v_4 & = & \rho_2+6/5\rho_3 = 0.
\end{eqnarray*}

The solution of this system of equations is $\rho_1=0$,
$\rho_2=-121/(20k)$ and $\rho_3=-467/(96k)$. Thus the
corresponding solution obtained after replacing these values of
$\rho_1,\rho_2$ and $\rho_3$ into (\ref{e105}), is unique and
equal to:
$$
\left(\begin{array}{c}
 u_{1} \cr  v_{1} \cr  u_{2} \cr  v_{2} \cr  u_{3} \cr  v_{3} \cr  u_{4} \cr  v_{4} \cr
\end{array}\right)=\fraca{1}{k}\left(\begin{array}{c}
 0 \cr  0 \cr  -9/10 \cr  -897/80 \cr  39/10 \cr  -121/120 \cr  0 \cr  -951/80 \cr
\end{array}\right).
$$

Note that the uniqueness of solution implies that the imposed
boundary conditions are enough to avoid rigid solid movements and
therefore the supports would be able to react to external forces
acting on the structure in such a way that the compatibility
conditions (equilibrium) hold. The solution is unique because for
the submatrix corresponding to $\rho$ coefficients and $u_1,v_1$
and $v_4$ in (\ref{e105}), we have:
  $$\mbox{Rank}\left(\begin{matrix}1 & 0 & 4/5 \cr 0 & 1 & -3/5 \cr 0 & 1 & 6/5 \cr\end{matrix}\right)=3\quad
   \Leftrightarrow \quad
\mbox{rank}\left(\begin{array}{c}
  {\bf K} \\
 - \\
  {\bf B} \\
\end{array}\right)=2m=8.
$$
}
\end{example}

\begin{example}[An isostatic truss structure]\label{ex31}{\em
Consider the structure shown in Figure \ref{f15} subject to three
forces $P_1, P_2$ and $P_3$. Applying the superposition principle
and assuming that the forces acting on the structure are: \begin{eqnarray*}
   {\bf F}_1 & = & P_1\left(\begin{matrix}0 & 0 & 0 & 0 & 1 & 0 & -1 & 0 \cr
\end{matrix}\right)^T,\\
  {\bf F}_2 & = & P_2\left(\begin{matrix}3/4 & 1 & 0 & 0 & 0
& -1 & -3/4 & 0 \cr \end{matrix}\right)^T \\
 {\bf F}_3 & = & P_3\left(\begin{matrix}3/2 & 1 & 0 & -1 & 0
& 0 & -3/2 & 0 \cr \end{matrix}\right)^T \\
{\bf F}_4 & = & P_4\left(\begin{matrix}9/4 & 1 & 0 & 0 & 0 & 0 & -9/4 &
-1 \cr \end{matrix}\right)^T
\end{eqnarray*}
that satisfy the compatibility conditions
(\ref{equ13})-(\ref{equ15}), the associated system of equations
leads to the general solution:
\begin{equation}\label{equ29}
\renewcommand{\arraycolsep}{0.00cm}
\left(\begin{array}{c}
 u_{1} \cr  v_{1} \cr  u_{2} \cr  v_{2} \cr  u_{3} \cr  v_{3} \cr  u_{4} \cr  v_{4} \cr
\end{array}\right)\!\!=\!\!\renewcommand{\arraycolsep}{-0.09cm}\fraca{P_1}{k}\!\!\left(\renewcommand{\arraystretch}{1.7}\begin{array}{c}
 \fraca{3}{5} \cr  \fraca{9}{20} \cr  \fraca{3}{5} \cr  -\fraca{9}{20} \cr  \fraca{6}{5} \cr  0 \cr  0 \cr  0 \cr
\end{array}\right)\!\!+\!\!\fraca{P_2}{k}\!\!\left(\renewcommand{\arraystretch}{1.7}\begin{array}{c}
 \fraca{9}{20} \cr  \fraca{19}{10} \cr  \fraca{9}{20} \cr  -\fraca{27}{80} \cr  \fraca{9}{10} \cr  0 \cr  0 \cr  0 \cr
\end{array}\right)\!\!+\!\!\fraca{P_3}{k}\!\!\left(\renewcommand{\arraystretch}{1.7}\begin{array}{c}
 \fraca{341}{210} \cr  \fraca{25}{32} \cr  \fraca{233}{120} \cr  -\fraca{233}{160} \cr  \fraca{9}{5} \cr  0 \cr  0 \cr  0 \cr
\end{array}\right)\!\!+\!\!\fraca{P_4}{k}\!\!\left(\renewcommand{\arraystretch}{1.7}\begin{array}{c}
 \fraca{287}{60} \cr  -\fraca{27}{40} \cr  \fraca{179}{60} \cr  -\fraca{27}{40} \cr  \fraca{9}{5} \cr  0 \cr  0 \cr  0 \cr
\end{array}\right)+\renewcommand{\arraycolsep}{0.05cm}\rho_1\!\!\left(\begin{array}{c}
1 \cr  0 \cr  1 \cr  0 \cr  1 \cr  0 \cr  1 \cr 0 \cr
\end{array}\right)+\rho_2\!\!\left(\begin{array}{c}
0 \cr  1 \cr  0 \cr  1 \cr  0 \cr  1 \cr  0 \cr 1 \cr
\end{array}\right)+\rho_3\!\!\left(\renewcommand{\arraystretch}{1.7}\begin{array}{c}
\fraca{4}{5} \cr  -\fraca{3}{5} \cr  \fraca{4}{5} \cr  \fraca{3}{5} \cr  0 \cr  0 \cr  0 \cr \fraca{6}{5 }\cr
\end{array}\right),
\renewcommand{\arraystretch}{1.0}
\end{equation}
where $P_1, P_2, P_3,P_4,\rho_1,\rho_1,\rho_3\in\R$. From a
mathematical point of view this is a linear space, i.e., the
solutions are the linear combinations of seven given vectors.

\begin{figure}[h]
\begin{center}
\includegraphics*[scale=0.50]{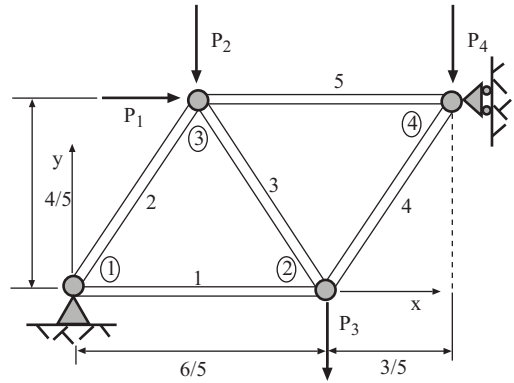}
\caption{Isostatic structure in Example \ref{ex31} showing the
applied loads.}\label{f15}
\end{center}
\end{figure}

}
\end{example}

\begin{example}[A critical truss structure]\label{ex30}{\em
If we consider the same structure as in Example \ref{ex32} with
other boundary conditions, for example, $u_1=0, v_1=0$ and
$u_2=0$, then from (\ref{e105}) we have:
\begin{eqnarray*}
    u_1 & = & 467/(120k)+\rho_1+4/5\rho_3 = 0  \\
    v_1 & = & 501/(160k)+\rho_2-3/5\rho_3 = 0  \\
    u_2 & = & 359/(120k) +\rho_1+4/5\rho_3 = 0,
\end{eqnarray*}
which is an incompatible system of equations, because for the
  submatrix associated with the $\rho$'s and $u_1,v_1$ and
  $u_4$ in (\ref{e105}) we have
  $$\mbox{Rank}\left(\begin{matrix}1 & 0 & 4/5 \cr 0 & 1 & -3/5 \cr 1 & 0 & 4/5 \cr\end{matrix}\right)=2<3.$$

 Adding the equations imposed by the boundary
  conditions to the system (\ref{e100}), considering the
  external forces (see Figure \ref{f7}) ${\bf F}=1/8\left(\begin{matrix}9 & 12& 0& -8& 8& -8& -17& 4\cr\end{matrix}\right)^T$
  that satisfy the
compatibility conditions
 (\ref{equ13})-(\ref{equ15})(equilibrium conditions) and applying the
orthogonalization algorithm the following set of all possible
solutions is obtained:
\begin{equation}\label{equ16}
\left(\begin{array}{c}
 u_{1} \cr  v_{1} \cr  u_{2} \cr  v_{2} \cr  u_{3} \cr  v_{3} \cr  u_{4} \cr  v_{4} \cr
\end{array}\right)=\fraca{1}{k}\left(\begin{array}{c}
 0 \cr  0 \cr  0 \cr  -49/16 \cr  -1/24 \cr  -37/16 \cr  -73/24 \cr  0 \cr
\end{array}\right)+\rho\left(\begin{array}{c}
0 \cr  0 \cr  0 \cr  6/5 \cr  -4/5 \cr  3/5 \cr  -4/5 \cr 9/5 \cr
\end{array}\right).
\end{equation}

\begin{figure}[h]
\begin{center}
\includegraphics*[scale=0.50]{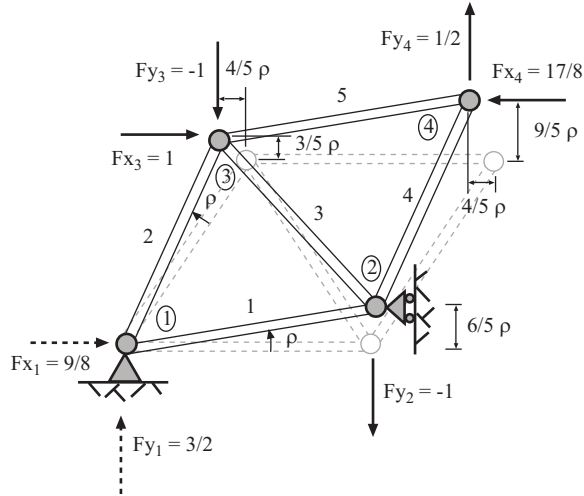}
\caption{Illustration of the set of general solutions associated
with the boundary conditions $u_1=0, v_1=0$ and $u_2=0$ for the critical
two-dimensional truss example.}\label{f7}
\end{center}
\end{figure}

From an engineering point of view, this solution must be
interpreted as follows:
\begin{enumerate}
  \item The first vector is a particular solution, i.e., a
  solution to the stated problem. Note that it satisfies equation
  (\ref{e100}) for the ${\bf F}$ values considered and the boundary conditions $u_1=0,
v_1=0$ and $u_2=0$.
  \item The second vector corresponds to a solution of the
  associated homogeneous problem, and represents the rotation of
  the structure as a rigid solid with respect node $1$, as it is shown in Figure
  \ref{f7}. This implies that there is a set of infinite solutions (rotations), and
  that the actual boundary conditions are not able to avoid the
  rigid solid displacement of the structure without the development of new forces in the bars under the hypothesis of small displacements.
  Moreover, the
  supports are not capable of supplying the reactions required to satisfy the equilibrium conditions
  for all possible forces acting on the nodes of the structure.
\end{enumerate}

\begin{figure}[h]
\begin{center}
\includegraphics*[scale=0.50]{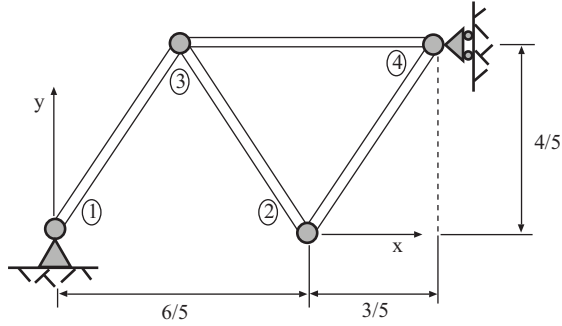}
\caption{The mechanism structure analyzed in Example
\ref{ex25}.}\label{f8}
\end{center}
\end{figure}

The same result is obtained considering
$$
{\bf B}=\left(%
\begin{array}{cccccccc}
  1 & 0 & 0 & 0 & 0 & 0 & 0 & 0 \\
  0 & 1 & 0 & 0 & 0 & 0 & 0 & 0 \\
  0 & 0 & 1 & 0 & 0 & 0 & 0 & 0 \\
\end{array}%
\right)\quad \Leftrightarrow \quad
\mbox{rank}\left(\begin{array}{c}
  {\bf K} \\
 - \\
  {\bf B} \\
\end{array}\right)=2m-1=7
$$
that implies that there is one rigid solid rotation ($\rho$)
allowed.
 }
\end{example}

\begin{example}[A Mechanism structure]\label{ex25}{\em
Consider the simple mechanism truss structure shown in Figure
\ref{f8}. The system of equations resulting from assembling the
global element stiffness matrices becomes:
\begin{equation}\label{equ17}
{\bf K}{\bf
u}=\fraca{k}{150}\left(\renewcommand{\arraycolsep}{0.03cm}\renewcommand{\arraystretch}{1.3}\begin{array}{cccccccc}
54 & 72 & 0 & 0 & -54 & -72 & 0 & 0\\
72 & 96 & 0 & 0 & -72 & -96 & 0 & 0\\
0 & 0 & 108 & 0 & -54 & 72 & -54 & -72\\
0 & 0 & 0 & 192 & 72 & -96 & -72 & -96\\
-54 & -72 & -54 & 72 & 233 & 0 & -125 & 0\\
-72 & -96 & 72 & -96 & 0 & 192 & 0 & 0\\
0 & 0 & -54 & -72 & -125 & 0 & 179 & 72\\
0 & 0 & -72 & -96 & 0 & 0 & 72 & 96
\end{array}\right)\left(\begin{array}{c}
 u_{1} \cr  u_{2} \cr  u_{3} \cr  u_{4} \cr  u_{5} \cr  u_{6} \cr  u_{7} \cr  u_{8} \cr
\end{array}\right)=\left(\begin{array}{c}
F_{x_1} \cr F_{y_1} \cr F_{x_2}  \cr F_{y_2}  \cr F_{x_3}  \cr
F_{y_3} \cr F_{x_4}  \cr F_{y_4}  \cr
\end{array}\right)\end{equation}
\renewcommand{\arraystretch}{1}
and the corresponding compatibility conditions, using the method
described in Section \ref{s3.1}, become:
 \begin{eqnarray}
 0 &=& F_{x_1}+F_{x_2}+F_{x_3}+F_{x_4} \label{equ18}\\
 0 &=&   F_{y_1}+F_{y_2}+F_{y_3}+F_{y_4} \label{equ19}\\
 0 &=&   4/5 F_{x_1}-3/5 F_{y_1}\label{equ20}\\
 0 &=&   4/5 F_{x_2}+3/5 F_{y_2}+6/5 F_{y_4},\label{equ21}
 \end{eqnarray}
where (\ref{equ18}) and (\ref{equ19}) express the equilibrium of
horizontal and vertical forces, respectively, and (\ref{equ20})
and (\ref{equ21}) establish the equilibrium of moments with
respect to node $3$ of the left and right substructures,
respectively.

Assuming that the forces acting on the structure satisfy the
compatibility conditions (\ref{equ18})-(\ref{equ21}) and take
values: ${\bf F}=1/8\left(\begin{matrix}9 & 12 & 0 & -8 & 8
& -8 & -17 & 4 \end{matrix}\right)^T,$
 the system (\ref{equ17}) leads to the general solution:
\begin{equation}\label{equ22c}
\left(\begin{array}{c}
 u_{1} \cr  v_{1} \cr  u_{2} \cr  v_{2} \cr  u_{3} \cr  v_{3} \cr  u_{4} \cr  v_{4} \cr
\end{array}\right)=\fraca{1}{k}\left(\begin{array}{c}
 49/8 \cr  0 \cr  3/2 \cr  -61/32 \cr  3 \cr  0 \cr  0 \cr  0 \cr
\end{array}\right)+\rho_1\left(\begin{array}{c}
1 \cr  0 \cr  1 \cr  0 \cr  1 \cr  0 \cr  1 \cr 0 \cr
\end{array}\right)+\rho_2\left(\begin{array}{c}
0 \cr  1 \cr  0 \cr  1 \cr  0 \cr  1 \cr  0 \cr 1 \cr
\end{array}\right)+\rho_3\left(\begin{array}{c}
4/5 \cr  -3/5 \cr  0 \cr  0 \cr  0 \cr  0 \cr  0 \cr 0 \cr
\end{array}\right)+\rho_4\left(\begin{array}{c}
0 \cr  0 \cr  4/5 \cr  3/5 \cr  0 \cr  0 \cr  0 \cr 6/5 \cr
\end{array}\right),
\end{equation}
where $\rho_1,\rho_1,\rho_3,\rho_4\in\R$. From a mathematical
point of view this is an affine space, i.e., the sum of a given
vector (the first one) plus a linear space of dimension $4$ (an
arbitrary linear combination of 4 linearly independent vectors).
The graphical interpretation of these four vectors is shown in
Figures \ref{f9} (a-d), i.e., a horizontal translation, a
vertical translation and two independent rotations with respect to
node 3. Note that eliminating the bar $1$-$2$ allows the bar
$1$-$3$ and the right substructure to rotate with respect to node
$3$ independently. That is why there are two rotation parameters
$\rho_3$ and $\rho_4$. Note that in the solution (\ref{e105})
corresponding to the structure with this bar, the rotation vector
is the sum of the last two vectors in (\ref{equ22c}).

\begin{figure}[h]
\begin{center}
\includegraphics*[scale=0.60]{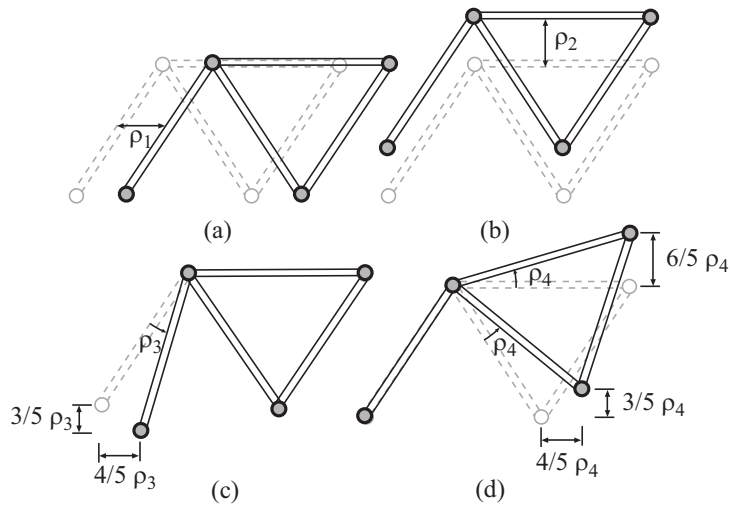}
\caption{Illustration of the four basic vectors of the linear
space component of the solution corresponding to Example
\ref{ex25}. They correspond to a horizontal translation, a
vertical translation and two independent rotations with respect to
node 3.}\label{f9}
\end{center}
\end{figure}

If we add the boundary conditions $u_1=0, v_1=0$ and $u_4=0$, then
from (\ref{equ22c}) we obtain the system of linear equations:
\begin{eqnarray*}
    u_1 & = & 49/(8k)+\rho_1+4/5\rho_3 = 0  \\
    v_1 & = & \rho_2-3/5\rho_3 = 0  \\
    u_4 & = & \rho_1 = 0
\end{eqnarray*}
which solution is $\rho_1=0$, $\rho_2=-147/(32k)$ and
$\rho_3=-245/(32k)$, but $\rho_4$ becomes free.

Considering the boundary conditions matrix
$$
{\bf B}=\left(%
\begin{array}{cccccccc}
  1 & 0 & 0 & 0 & 0 & 0 & 0 & 0 \\
  0 & 1 & 0 & 0 & 0 & 0 & 0 & 0 \\
  0 & 0 & 0 & 0 & 0 & 0 & 1 & 0 \\
\end{array}%
\right)\quad \Leftrightarrow \quad
\mbox{rank}\left(\begin{array}{c}
  {\bf K} \\
 - \\
  {\bf B} \\
\end{array}\right)=2m-1=7
$$
that implies that there is $1$ rigid solid movements allowed under
the assumption of small displacements without strains.

Then, the solution in this case becomes:
$$
\left(\begin{array}{c}
 u_{1} \cr  v_{1} \cr  u_{2} \cr  v_{2} \cr  u_{3} \cr  v_{3} \cr  u_{4} \cr  v_{4} \cr
\end{array}\right)=\fraca{1}{k}\left(\begin{array}{c}
 0 \cr  0 \cr  -3/2 \cr  -13/2 \cr  3 \cr  -147/32 \cr  0 \cr  -147/32 \cr
\end{array}\right)+\rho_4\left(\begin{array}{c}
0 \cr  0 \cr  4/5 \cr  3/5 \cr  0 \cr  0 \cr  0 \cr 6/5 \cr
\end{array}\right),
$$
that implies that the actual boundary conditions do not prevent
the rotation of the right substructure with respect node $3$ as
shown in Figure \ref{f10}.

\begin{figure}[h]
\begin{center}
\includegraphics*[scale=0.50]{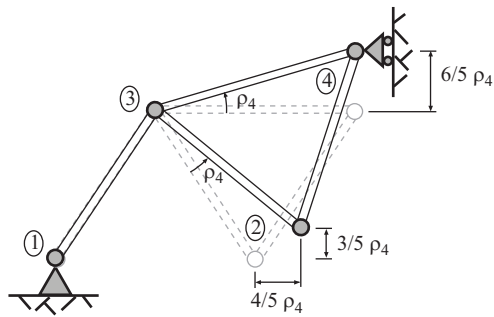}
\caption{Illustration of how the boundary conditions $u_1=0,
v_1=0$ and $u_4=0$ do not prevent the rotation of the right
substructure with respect node $3$ in a mechanism structure.}\label{f10}
\end{center}
\end{figure}
}
\end{example}

\begin{figure}[h]
\begin{center}
\includegraphics*[scale=0.50]{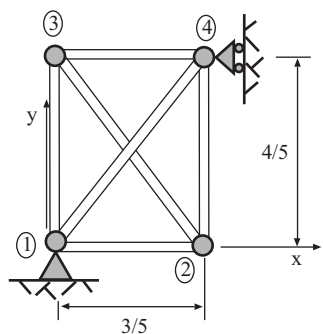}
\caption{The hyperstatic structure analyzed in Example
\ref{ex26}.}\label{f11}
\end{center}
\end{figure}

\begin{example}[An hyperstatic structure]\label{ex26}{\em
Consider now the structure given in Figure \ref{f11} where an
additional bar joining nodes $1$ and $4$ has been added. The
system of equations resulting from assembling the global element
stiffness matrices in this case is:
\begin{equation}\label{equ23}
{\bf K}{\bf
u}=\fraca{k}{300}\left(\renewcommand{\arraycolsep}{0.03cm}\renewcommand{\arraystretch}{1.3}\begin{array}{cccccccc}
608 & 144 & -500 & 0 & 0 & 0 & -108 & -144\\
144 & 567 & 0 & 0 & 0 & -375 & -144 & -192\\
-500 & 0 & 608 & -144 & -108 & 144 & 0 & 0\\
0 & 0 & -144 & 567 & 144 & -192 & 0 & -375\\
0 & 0 & -108 & 144 & 608 & -144 & -500 & 0\\
0 & -375 & 144 & -192 & -144 & 567 & 0 & 0\\
-108 & -144 & 0 & 0 & -500 & 0 & 608 & 144\\
-144 & -192 & 0 & -375 & 0 & 0 & 144 & 567
\end{array}\right)\left(\begin{array}{c}
 u_{1} \cr  u_{2} \cr  u_{3} \cr  u_{4} \cr  u_{5} \cr  u_{6} \cr  u_{7} \cr  u_{8} \cr
\end{array}\right)=\left(\begin{array}{c}
F_{x_1} \cr F_{y_1} \cr F_{x_2}  \cr F_{y_2}  \cr F_{x_3}  \cr
F_{y_3} \cr F_{x_4}  \cr F_{y_4}  \cr
\end{array}\right)\end{equation}
\renewcommand{\arraystretch}{1} and the compatibility conditions become:
 \begin{eqnarray}
 0 &=& F_{x_1}+F_{x_2}+F_{x_3}+F_{x_4} \label{equ24}\\
 0 &=&   F_{y_1}+F_{y_2}+F_{y_3}+F_{y_4} \label{equ25}\\
 0 &=&   4/5 F_{x_1}+4/5 F_{x_2}+3/5 F_{y_2}+3/5 F_{y_4},\label{equ26}
 \end{eqnarray}
which correspond to the equilibrium of horizontal forces, vertical
forces and moments with respect to node $3$.

Assuming that the forces acting on the structure satisfy the
compatibility conditions (\ref{equ24})-(\ref{equ26}) and take
values ${\bf F}=1/4\left(\begin{matrix}3 & 8 & 0 & -4 & 4
& -4 & -7 & 0 \end{matrix}\right)^T,$ the following general solution is obtained:
\begin{equation}\label{equ22b}
\left(\begin{array}{c}
 u_{1} \cr  v_{1} \cr  u_{2} \cr  v_{2} \cr  u_{3} \cr  v_{3} \cr  u_{4} \cr  v_{4} \cr
\end{array}\right)=\fraca{1}{k}\left(\begin{array}{c}
 353/960 \cr  521/540 \cr  11/40 \cr  -343/540 \cr  133/192 \cr  0 \cr  0 \cr  0 \cr
\end{array}\right)+\rho_1\left(\begin{array}{c}
1 \cr  0 \cr  1 \cr  0 \cr  1 \cr  0 \cr  1 \cr 0 \cr
\end{array}\right)+\rho_2\left(\begin{array}{c}
0 \cr  1 \cr  0 \cr  1 \cr  0 \cr  1 \cr  0 \cr 1 \cr
\end{array}\right)+\rho_3\left(\begin{array}{c}
4/5 \cr  0 \cr  4/5  \cr  3/5  \cr  0 \cr  0 \cr  0 \cr 3/5  \cr
\end{array}\right);\quad \rho_1,\rho_1,\rho_3\in\R.
\end{equation}
From a mathematical point of view it is an affine linear space,
i.e., the sum of a given vector (the first one) plus a linear
space of dimension $3$ (an arbitrary linear combination of 3
linearly independent vectors).

\begin{figure}[h]
\begin{center}
\includegraphics*[scale=0.50]{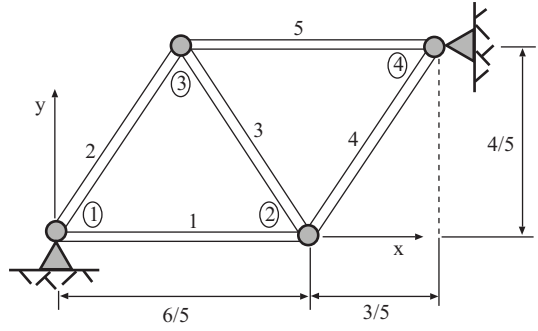}
\caption{Hyperstatic truss structure of Example \ref{ex27}.}\label{f12}
\end{center}
\end{figure}
}
\end{example}

\begin{example}[Hyperstatic structure]\label{ex27}{\em Consider the particular
truss structure in Figure \ref{f12}, subject to $3$ external
forces. Note that it is the same example as the initial one but
adding a new boundary condition ($v_4=0$), thus we add a new
equation to the system (\ref{e100}) but we have a new unknown
variable, the vertical support reaction on node $4$ ($F_{y_4}$).
Applying the superposition principle we transform our problem as
it is shown in Figure \ref{f13}.

The set off all possible solutions will be composed by
(\ref{e105}) and the particular solution associated with the
vector
\begin{equation}\label{equ24a}{\bf F}=\fraca{1}{4}\left(\begin{matrix}-9 & -4 & 0 & 0 & 0
& 0 & 9 & 4 \end{matrix}\right)^T,
\end{equation}
 which is
\begin{equation}\label{equ2a}
\left(\begin{array}{c}
 u_{1} \cr  v_{1} \cr  u_{2} \cr  v_{2} \cr  u_{3} \cr  v_{3} \cr  u_{4} \cr  v_{4} \cr
\end{array}\right)=\fraca{1}{k}\left(\begin{array}{c}
 467/120 \cr  501/160 \cr  359/120 \cr  -359/160 \cr  39/10 \cr  0 \cr  0 \cr  0 \cr
\end{array}\right)+\fraca{F_{y_4}}{k}\left(\begin{array}{c}
 -287/60 \cr  27/40 \cr  -179/60 \cr  27/40 \cr  -9/5 \cr  0 \cr  0 \cr  0 \cr
\end{array}\right)+\rho_1\left(\begin{array}{c}
1 \cr  0 \cr  1 \cr  0 \cr  1 \cr  0 \cr  1 \cr 0 \cr
\end{array}\right)+\rho_2\left(\begin{array}{c}
0 \cr  1 \cr  0 \cr  1 \cr  0 \cr  1 \cr  0 \cr 1 \cr
\end{array}\right)+\rho_3\left(\begin{array}{c}
4/5 \cr  -3/5 \cr  4/5 \cr  3/5 \cr  0 \cr  0 \cr  0 \cr 6/5 \cr
\end{array}\right),
\end{equation}
where $F_{y_4},\rho_1,\rho_2,\rho_3\in\R$, that from a
mathematical point of view is an affine linear space, i.e., the
sum of a given vector (the first one) plus a linear space of
dimension $4$ (an arbitrary linear combination of 4 linearly
independent vectors). The graphical interpretation of the last
three vectors is shown in Figure \ref{f3}, whereas the first one
is the particular solution associated with the system of forces
(\ref{equ24a}).

\begin{figure}[h]
\begin{center}
\includegraphics*[width=\textwidth]{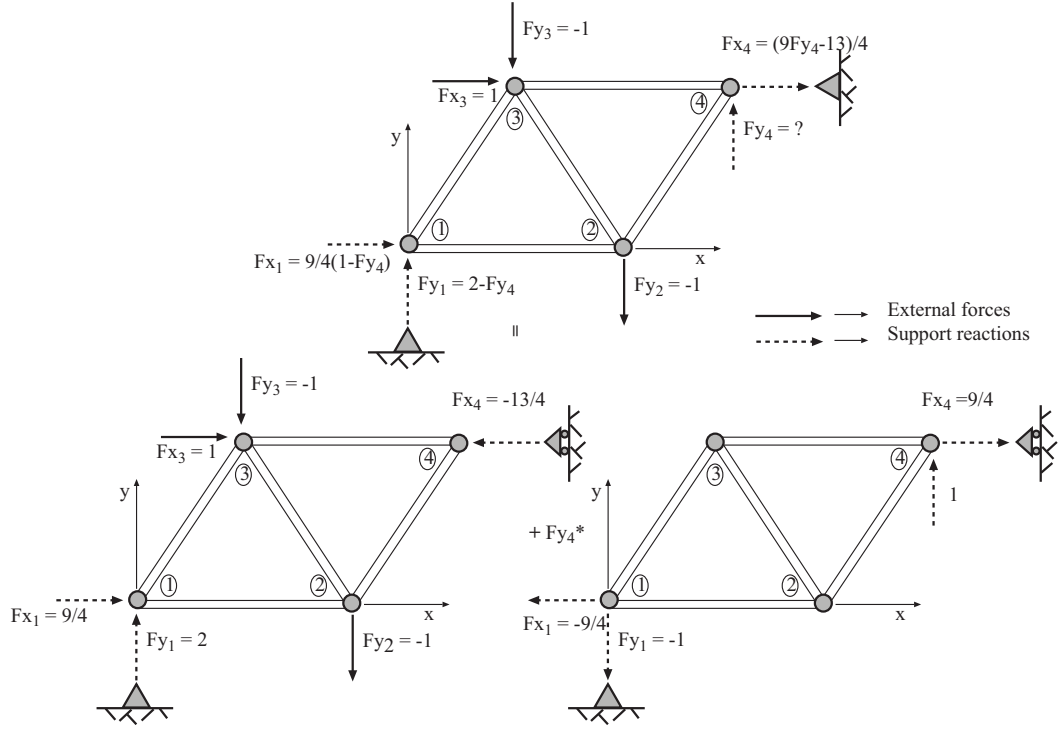}
\caption{Illustration of the superposition principle applied to
Example \ref{ex27}.}\label{f13}
\end{center}
\end{figure}

If we add the boundary conditions $u_1=0, v_1=0, u_4=0$ and
$v_4=0$, then from (\ref{equ23}) we have:

\begin{equation}\label{equ25a}
\left(\begin{array}{c}
 u_{1}-467/(120k) \cr  v_{1}-501/(160k) \cr u_{4} \cr  v_{4} \cr
\end{array}\right)=\left(\begin{array}{c}
 -287/(60k) \cr  27/(40k) \cr 0 \cr  0 \cr
\end{array}\begin{array}{c}
1 \cr  0 \cr  1 \cr 0 \cr
\end{array}\begin{array}{c}
0 \cr  1 \cr   0 \cr 1 \cr
\end{array}\begin{array}{c}
4/5 \cr  -3/5 \cr  0 \cr 6/5 \cr
\end{array}\right)
\left(\begin{array}{c} F_{y_4} \cr  \rho_1 \cr \rho_2 \cr \rho_3
\cr
\end{array}\right),
\end{equation}
which solution is $F_{y_4}=317/269$, $\rho_1=0$,
$\rho_2=-11267/(4304k)$ and $\rho_3=56335/(25824k)$. Thus the
solution is unique and equal to:
$$
\left(\begin{array}{c}
 u_{1} \cr  v_{1} \cr  u_{2} \cr  v_{2} \cr  u_{3} \cr  v_{3} \cr  u_{4} \cr  v_{4} \cr
\end{array}\right)=\fraca{1}{k}\left(\begin{array}{c}
 0 \cr  0 \cr  657/538 \cr  -11867/4304 \cr  957/538 \cr  -11267/4304 \cr  0 \cr  0 \cr
\end{array}\right).
$$
}
\end{example}

\section{Solving Systems of Inequalities}\label{s6}In many engineering applications we
find systems of linear inequalities of the form:
%
%\begin{equation}\label{eq1r}
%\begin{array}{ccccccc}
%a_{11}x_1 & +  a_{12}x_2 & +  \cdots  & +  a_{1n}x_n & \le & b_1,\\
%a_{21}x_1 & +  a_{22}x_2 & +  \cdots  & +  a_{2n}x_n & \le & b_2,\\
%\cdots  & \cdots    & \cdots  &  \cdots   &   & \cdots \\
%a_{m1}x_1 & +  a_{m2}x_2 & +  \cdots  & +  a_{mn}x_n & \le & b_m,
%\end{array}
%\end{equation}
%
\begin{equation}\label{eq1r}
{\bf A}{\bf x}\le {\bf b}.
\end{equation}

 Before trying to solve a system of
inequalities of this type it is interesting to check whether or
not the system is compatible, that is, if it has some solutions.
Even, in many cases we can be interested in obtaining the
conditions for the system to be compatible in terms of the
independent terms $b_1,b_2,\dots,b_m$. These conditions always
have a physical or engineering meaning that raises some light
about the problem under consideration. Once the system is proven to have solution we solve the system and
obtain all its possible solutions.

To solve these two problems the
$\Gamma$ algorithm is used (see Jubete \cite{Jubete:91,Jubete:93},
Padberg \cite{Padberg:95}, Castillo, Jubete, Pruneda and Solares
\cite{CastilloJPS:00a}, Castillo, Esquivel y Pruneda
\cite{CastilloEP:01} and Castillo, Conejo, Pedregal, Garc\'{\i}a
and Alguacil \cite{CastilloCPGA:01}),  that gives the dual cone of
a given cone and is the key tool to discuss the compatibility
problem and solve the system of inequalities.

Since the concepts of cone and dual cone are used, we start with
their definitions.

\begin{definition}[Polyhedral convex cone]\index{Polyhedral convex cone!definition}\index{Definition!polyhedral convex
cone}Let ${\bf A}$ be a matrix, and $\{ {\bf a}_1,\dots,{\bf
a}_m\}$ be its column vectors. The set
\begin{displaymath}
{\bf A}_\pi\equiv \left\{ {\bf x}\in \R^n\;|\;{\bf x}=\pi_1{\bf
a}_1+\dots+\pi_m{\bf a}_m\;\;\;\mbox{with}\;\;\;\pi_j\geq
0;j=1,\dots ,m \right\}
\end{displaymath}
of all nonnegative linear combinations of the column vectors of
${\bf A}$ is known as the polyhedral convex cone generated by
${\bf a}_1,\dots ,{\bf a}_m$ (its generators), and is denoted
${\bf A}_\pi$.
\end{definition}

A cone ${\bf A}_\pi$ can be written as the sum of a linear space
${\bf V}_\rho$ plus a pure (acute) cone ${\bf W}_\pi$, i.e., ${\bf
A}_\pi={\bf V}_\rho+{\bf W}_\pi$.

In this paper we use the Greek letter $\pi$ to refer to
non-negative real numbers.

\begin{definition}[Nonpositive dual or polar cone]Let ${\bf A}_\pi$ be a cone in $\R^n$ with
generators ${\bf a}_1,\dots,{\bf a}_k$. The nonpositive dual of
${\bf A}_\pi$, denoted ${\bf A}^p_\pi$, is defined as the set
\begin{displaymath}
{\bf A}^p_\pi\equiv \left\{ {\bf u}\in \R^n\;|\;{\bf A}^T{\bf
u}\leq {\bf 0} \right\} \equiv \left\{ {\bf u}\in \R^n\;|\;{\bf
a}_i^T {\bf u}\leq 0;\;i=1,\dots, k \right\}
\end{displaymath}
that is, the set of all vectors such that their dot products by
all vectors in ${\bf A}_\pi$ are nonpositive.
\end{definition}

\begin{definition}[Polytope]Let ${\bf A}$ be a matrix, and $\{ {\bf a}_1,\dots,{\bf
a}_m\}$ be its column vectors. The set
\begin{displaymath}
{\bf A}_\lambda\equiv \left\{ {\bf x}\in \R^n\;|\;{\bf
x}=\lambda_1{\bf a}_1+\dots+\lambda_m{\bf
a}_m\;\;\;\mbox{with}\;\;\;\lambda_j\geq 0;j=1,\dots
,m;\sum\limits_{i=1}^m \lambda_i=1 \right\}
\end{displaymath}
of all linear convex combinations of the column vectors of ${\bf
A}$ is known as the polytope generated by ${\bf a}_1,\dots ,{\bf
a}_m$ (its generators), and is denoted ${\bf A}_\lambda$.
\end{definition}

\subsection{Deciding whether or not a system of linear inequalities is compatible}\label{s5.10}
In this section we show how to analyze the compatibility of a
system of linear inequalities.

First, we discuss the compatibility of a particular system of the
form:
\begin{equation}\label{eq1p}
\begin{array}{cccccccc}
a_{11}x_1 & +  a_{12}x_2 & + & \cdots  & +  a_{1n}x_n & = & b_1,\\
a_{21}x_1 & +  a_{22}x_2 & +  &\cdots  & +  a_{2n}x_n & = & b_2,\\
\cdots  & \cdots    & \cdots  &  \cdots   &   & \cdots \\
a_{m1}x_1 & +  a_{m2}x_2 & + & \cdots  & +  a_{mn}x_n & = & b_m\\
\multicolumn{5}{r}{x_1,x_2,\cdots,x_n} & \ge & 0
\end{array}
\end{equation}
that can be written as
\begin{eqnarray}\label{eq4a}
x_1 \left(\begin{array}{cccc}
 a_{11}\\ a_{21}\\ \vdots \\a_{m1}
\end{array}\right)+
x_2 \left(\begin{array}{cccc}
 a_{12}\\ a_{22}\\ \vdots \\a_{m2}
\end{array}\right)+\cdots
+x_n \left(\begin{array}{cccc}
 a_{1n}\\ a_{2n}\\ \vdots \\a_{mn}
\end{array}\right) &=& \left(\begin{array}{cccc}
 b_1\\ b_2\\ \vdots \\b_m
\end{array}\right) \\
x_1,x_2,\cdots,x_n & \ge & 0\nonumber
\end{eqnarray}

Expression (\ref{eq4a}) shows that the given system is compatible
if and only if the vector ${\bf b}=(b_1,\dots ,b_m)^T$ belongs to
the cone generated by the set of column vectors $\{{\bf a}_1,{\bf
a}_2,\cdots,{\bf a}_n\}$ of the coefficient matrix ${\bf A}$,
i.e.,
\begin{equation}\label{e500}
{\bf b}\in {\bf A}_\pi={\bf b}\in \left({\bf A}_\pi^p\right)^p
\end{equation}

Thus, the compatibility problem reduces to finding the dual cone
${\bf V}_\rho+{\bf W}_\pi$ of the cone generated by the columns of
the coefficient matrix and checking that ${\bf b}^T{\bf V}={\bf
0}$ and ${\bf b}^T{\bf W}\le {\bf 0}$.

To analyze the compatibility of an arbitrary system of linear
inequalities, it can be converted to the case in (\ref{eq1p}),
using slack variables to convert the inequalities in equalities,
and one more artificial variable to convert the arbitrary
variables into no negative variables, that is, each variable $x_i$
can be converted to $x_i^*-x_0$, where $x_0,x_i^*\ge 0$.

\begin{example}[Compatibility of a linear system of equations in restricted variables]
\label{ex4}{\em Consider the following linear system:
\begin{equation}\label{e101}\begin{array}{r}
\begin{array}{rrrrcl}
& x_2 & - x_3 & - 2 x_4 & = & b_1 \\
& & x_3 & + x_4 & = &b_2 \\
& -x_2 & + x_3 & + 2 x_4 & = &b_3 \\
& x_2 & + x_3 & - x_4 & = & b_4\\
-x_1 & + 2 x_2 & + x_3 & + x_4 & = & b_5 \\
 \multicolumn{4}{r}{x_1,x_2,x_3,x_4} & \ge & 0.
\end{array}
\end{array}
\end{equation}
For it to be compatible, the vector ${\bf b}=(
b_1,b_2,b_3,b_4,b_5)^T$ must belong to the cone ${\bf C}_\pi$
generated by the columns of the coefficient matrix, that is to
say, it must belong to the dual of the dual of ${\bf C}_\pi$.
Thus, the compatibility problem reduces to finding the dual cone
${\bf C}_\pi^p\equiv{\bf V}_\rho+{\bf W}_\pi$  and checking that
${\bf b}^T{\bf V}={\bf 0}$ and ${\bf b}^T{\bf W}\le {\bf 0}$.

 Since ${\bf C}_\pi^p$ is the cone:
 $$
{\bf C}_\pi^p=\left(\begin{matrix} 1 \cr 0 \cr 1 \cr 0\cr 0
    \end{matrix}\right)_\rho+\left(
\begin{array}{rrrr}
  -2 & 1 & -1 & -4 \\ -3 & 1 & -2 &
    -7 \\ 0 & 0 & 0 & 0 \\ 1 &
    -1 & 1 & 2 \\ 0 & 0 & 0 & 1 \\  \end{array}\right)_\pi,
 $$
 we obtain the desired compatibility conditions:
\begin{equation}
\begin{array}{c}
\left(\begin{matrix}b_1 & b_2 & b_3 & b_4 &b_5  \end{matrix}\right)
\left(\renewcommand{\arraystretch}{1.2}
\begin{array}{rrrr}
  -2 & 1 & -1 & -4 \\ -3 & 1 & -2 &
    -7 \\ 0 & 0 & 0 & 0 \\ 1 &
    -1 & 1 & 2 \\ 0 & 0 & 0 & 1 \\  \end{array}\right)\le {\bf 0},\\
\\
 \left(\begin{matrix}b_1 & b_2 & b_3 & b_4 &b_5  \end{matrix}\right)  \left(\begin{matrix} 1 & 0 & 1 & 0& 0 \end{matrix}\right)^T=b_1+b_3 =0.
\end{array}\end{equation}

}
\end{example}

\subsection{Solving a homogeneous system of linear inequalities}\label{s5.6a}Consider the homogeneous system of
inequalities
%
%\begin{equation}\label{eq1b}
%\begin{array}{cccccc}
%a_{11}x_1 & +  a_{12}x_2 & +  \cdots  & +  a_{1n}x_n & \leq & 0,\\
%a_{21}x_1 & +  a_{22}x_2 & +  \cdots  & +  a_{2n}x_n & \leq  & 0,\\
%\cdots\cdots  & \cdots\cdots    & \cdots\cdots   &  \cdots\cdots   & \leq   & \cdots \\
%a_{m1}x_1 & +  a_{m2}x_2 & +  \cdots  & +  a_{mn}x_n & \leq  & 0\\
%\end{array}
%\end{equation}
%which can be written as
%\begin{equation}\label{eq5}
%\begin{array}{ccc}
% (a_{11}, \ldots ,a_{1n})  (x_1, \ldots , x_n)^T & \leq  & 0,\\
% (a_{21}, \ldots ,a_{2n})  (x_1, \ldots , x_n)^T & \leq  &0,\\
%  \cdots \cdots \cdots \cdots \cdots\cdots\cdots\cdots \cdots   & \leq  & \cdots\\
% (a_{m1}, \ldots ,a_{mn})  (x_1, \ldots , x_n)^T & \leq  & 0.\\
%\end{array}
%\end{equation}
%
\begin{equation}\label{eq1b}
{\bf A}{\bf x}\le {\bf 0}
\end{equation}
which can be written as
\begin{equation}\label{eq5}
{\bf a}^i{\bf x}^T\le 0;\;i=1,\ldots,m.
\end{equation}

Expression (\ref{eq5}) shows that $( x_1,\dots ,x_n)$ is the dual
cone of the row vectors $\{{\bf a}^1,{\bf a}^1,\dots,{\bf a}^m\}$
of ${\bf A}$.

Thus, obtaining the solution of the system (\ref{eq1b}) reduces to
determining the dual cone ${\bf A}^p_\pi$ of the cone generated by
the cone generated by the rows of matrix ${\bf A}$.

Thus, the general solution of an homogeneous system of linear
inequalities is a cone, that is, its general solution is of the
form
%
%$$
%\left(\begin{matrix}x_1\cr x_2\cr \vdots\cr  x_n\cr
%}\right)=\sum\limits_{i=1}^p \rho_i\left(\begin{matrix}v_{i1}\cr
%v_{i2}\cr \vdots\cr v_{in}\end{matrix}\right)+\sum\limits_{j=1}^q
%\pi_j\left(\begin{matrix}w_{j1}\cr w_{j2}\cr \vdots\cr w_{jn}\cr
%}\right);\quad \rho_i\in \R;\quad \pi_j\in \R^+.
%$$
%
$$
{\bf x}=\sum\limits_{i=1}^p \rho_i{\bf v}_i+\sum\limits_{j=1}^q
\pi_j{\bf w}_j;\quad \rho_i\in \R;\quad \pi_j\in \R^+.$$

\begin{example}[Solving an homogeneous system of linear inequalities]{\em
\label{ex5} Consider the system of equations
\begin{equation}\label{ex5a}\begin{array}{rrrrrrr}
& & & &-x_5 &\leq & 0\\
 x_1 &+ x_2 & &+ x_4 &- x_5&\leq & 0\\
 -x_1 &+ x_2 &- 2 x_3 &- x_4 &+ x_5&\leq & 0\\
 -2 x_1 & &+ x_3 &- x_4 &- x_5 &\leq & 0\\
2 x_1 &+ x_2 &- x_3 & &+ x_5&\leq & 0.\\
\end{array}
\end{equation}

To solve this system, we obtain the dual cone of the cone
generated by the rows coefficients and obtain the solution:
$$
\left(\begin{matrix}x_1\cr x_2\cr x_3\cr x_4\cr x_5\end{matrix}\right)\!=\!
\left(\renewcommand{\arraystretch}{1.2}\begin{array}{rrrrr}
\begin{matrix} 2 & 2 & 0 & -2 & -2 \cr -5 & -1 & -1 & -1 & 3 \cr -1 & 3
&
    -1 & -1 & 3 \cr -5 & -1 & 1 & 3 &3 \cr 0 & 0 & 0 & 0 & 4 \end{matrix}\end{array}\right) \left (\begin{array}{r}
\pi_1\\\pi_2\\\pi_3\\\pi_4\\\pi_5\\
\end{array}
\right);\quad \pi_i\in\R^+;\;i=1,2,\dots,5.
$$

}
\end{example}

\subsection{Solving a complete system of linear inequalities}\label{s5.7a}
Now consider the complete system of linear inequalities:
%
%\begin{equation}\label{eq1aaab}
%\begin{array}{ccccccc}
%a_{11}x_1 & +  a_{12}x_2 & +  \cdots  & +  a_{1n}x_n & \le  & b_1,\\
%a_{21}x_1 & +  a_{22}x_2 & +  \cdots  & +  a_{2n}x_n & \le  & b_2,\\
%\cdots  & \cdots    & \cdots  &  \cdots   &   & \cdots \\
%a_{m1}x_1 & +  a_{m2}x_2 & +  \cdots  & +  a_{mn}x_n & \le & b_m.
%\end{array}
%\end{equation}
%
\begin{equation}\label{eq1aaab}
{\bf A}{\bf x}\le {\bf b}
\end{equation}
where ${\bf A}\in \R^{m\times n}$ and ${\bf b}\in \R^{m}$.

Adding the artificial variable $x_{n+1}$, the constraint
$x_{n+1}=1$ and the redundant constraint $x_{n+1}\ge 0$ (it is a
key trick that allows the constraint $x_{n+1}=1$ to be easily
forced at the end of the process),  it can be written as
\begin{equation}\label{eq2}
\begin{array}{ccccrrc}
a_{11}x_1 & +  a_{12}x_2 & +  \cdots  & +  a_{1n}x_n & - b_1x_{n+1} & \le  & 0\\
a_{21}x_1 & +  a_{22}x_2 & +  \cdots  & +  a_{2n}x_n & - b_2x_{n+1}& \le  & 0\\
\cdots  & \cdots    & \cdots   &  \cdots   & \cdots    &   & \cdots \\
a_{m1}x_1 & +  a_{m2}x_2 & +  \cdots  & +  a_{mn}x_n & - b_mx_{n+1}& \le  & 0\\
a_{m1}x_1 & +  a_{m2}x_2 & +  \cdots  & +  a_{mn}x_n & - b_mx_{n+1}& \le  & 0\\
\multicolumn{5}{r}{-x_{n+1}} & \le & 0\\
\multicolumn{5}{r}{x_{n+1}} & = & 1.\\
\end{array}
\end{equation}

System (\ref{eq2}) can be written as
\begin{equation}\label{eq5a}
\begin{array}{rcc}
 (a_{11}, \ldots ,a_{1n},-b_1)  (x_1, \ldots , x_n, x_{n+1})^T & \le  & 0\\
 (a_{21}, \ldots ,a_{2n},-b_2)  (x_1, \ldots , x_n, x_{n+1})^T & \le  &0\\
  \cdots \cdots \cdots \cdots \cdots\cdots\cdots\cdots \cdots \cdots \cdots \ &  & \cdots\\
 (a_{m1}, \ldots ,a_{mn},-b_m)  (x_1, \ldots , x_n, x_{n+1})^T & \le  & 0\\
 -x_{n+1} & \le & 0\\
 x_{n+1} & = & 1.
\end{array}
\end{equation}

Expression (\ref{eq5a}) shows that $( x_1,\dots ,x_n, x_{n+1})$
belongs to the dual cone of the cone generated by the set of
vectors
$$\{( a_{11},\dots , a_{1n},-b_1),( a_{21},\dots , a_{2n},-b_2),
\dots , ( a_{m1},\dots , a_{mn},-b_m),(0,0,\cdots,0,-1)\}.$$

Then, it is clear that the solution of (\ref{eq2}) is the
intersection of that cone with the hyperplane $x_{n+1}=1$. Thus,
the solution of (\ref{eq1aaab}) is the projection on $X_1\times
\cdots \times X_n$ of the solution of (\ref{eq2}).  In other
words, the general solution of a complete system of linear
inequalities is a polyhedron, that is the sum of a linear space, a
cone and a polytope, that is, its general solution is of the form
%
%$$
%\left(\begin{matrix}x_1\cr x_2\cr \vdots\cr  x_n\cr
%}\right)=\sum\limits_{i=1}^p \rho_i\left(\begin{matrix}v_{i1}\cr
%v_{i2}\cr \vdots\cr v_{in}\end{matrix}\right)+\sum\limits_{j=1}^q
%\pi_j\left(\begin{matrix}w_{j1}\cr w_{j2}\cr \vdots\cr w_{jn}\cr
%}\right)+\sum\limits_{k=1}^r \lambda_k\left(\begin{matrix}u_{k1}\cr
%u_{k2}\cr \vdots\cr u_{kn}\end{matrix}\right);\quad \rho_i\in
%\R,\;\pi_j\in \R^+;\;\lambda_k\in
%\R^+;\sum\limits_{k=1}^r\lambda_k=1.
%$$
%
$$
{\bf x}=\sum\limits_{i=1}^p \rho_i{\bf v}_i+\sum\limits_{j=1}^q
\pi_j{\bf w}_i+\sum\limits_{k=1}^r \lambda_k{\bf u}_k;\quad \rho_i\in
\R,\;\pi_j\in \R^+;\;\lambda_k\in
\R^+;\sum\limits_{k=1}^r\lambda_k=1.
$$

\begin{example}[Solving a complete system of linear inequalities]{\em \label{ex6}
To solve the following system of inequalities:
\begin{equation}\label{e.1}\begin{array}{rrrrrr}
 x_1 &+ x_2 & &+ x_4 & \leq & 1\\
 -x_1 &+ x_2 &- 2 x_3 &- x_4 & \leq & -1\\
 -2 x_1 & &+ x_3 &- x_4   &\leq & 1\\
2 x_1 &+ x_2 &- x_3 & & \leq & -1,\\
\end{array}
\end{equation}
we use the auxiliary variable $x_5$ and the redundant constraint
$1=x_5\ge 0$. Then, the system (\ref{e.1}) can be written as:
\begin{equation}\begin{array}{rrrrrrr}
& & & &-x_5 &\leq & 0\\
 x_1 &+ x_2 & &+ x_4 &- x_5&\leq & 0\\
 -x_1 &+ x_2 &- 2 x_3 &- x_4 &+ x_5&\leq & 0\\
 -2 x_1 & &+ x_3 &- x_4 &- x_5 &\leq & 0\\
2 x_1 &+ x_2 &- x_3 & &+ x_5&\leq & 0\\
 & & &   &{x_5}  &= &1. \\
\end{array}
\end{equation}

Since the upper part is an homogeneous system, one need to find
the dual cone of the cone generated by the row coefficients. After
imposing condition $x_5=1$ one gets the solution:
\begin{equation}\begin{array}{rcl}
\left(\begin{matrix}x_1 \cr x_2 \cr x_3 \cr  x_4 \end{matrix}\right) & = &
\left ( \begin{array}{rrr}
-1/2  \\
3/4 \\
3/4  \\
3/4\\
\end{array}\right)+\left(
\begin{array}{rrrrrrr}
 2 & 2 & 0 & -2   \\
 -5 & -1& -1 & -1   \\
 -1 &  3 & -1 & -1  \\
-5 & -1 & 1 & 3  \\
\end{array}\right) \left (\begin{array}{r}
\pi_1\\\pi_2\\\pi_3\\\pi_4 \\
\end{array}
\right),
\end{array}
\end{equation}
which has no linear space part and whose polytope part reduce to a
single vector.
 }
\end{example}

\section{The Truss Analysis Problem Revisited}\label{s7}In this section we analyze the truss analysis problem, but
assuming additional constraints related to node displacements and
maximum bar compressions.

\begin{figure}[h]
\begin{center}
\includegraphics*[scale=0.50]{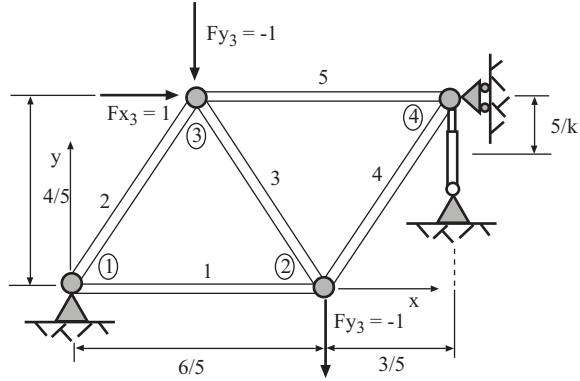}
\caption{Truss structure of Example \ref{ex28} showing the acting
forces and the support conditions.}\label{f14}
\end{center}
\end{figure}

\begin{example}[The lifting jack structure]\label{ex28}{\em
Consider the structure shown in Figure \ref{f14} where there is a
lifting jack that push the node $4$ in the vertical direction. We
want to obtain the minimum force that has to be applied by the
lifting jack, for the descent of node $4$ to be smaller than
$5/k$. We would like to get the set of all possible node
displacements for all the valid values of the force exerted by the
lifting jack.

As the boundary conditions in this case are $u_1=0$, $v_1=0$,
$u_4=0$ and $v_4\ge -5/k$, from (\ref{equ2a}) we can get the
following system of linear inequalities:
\begin{equation}\label{e200}
\begin{array}{rcl}
u_{1}-467/(120k) &  = &  -287/(60k) F_{y_4} +\rho_1+4/5\rho_3\\
 v_{1}-501/(160k) & = &  27/(40k) F_{y_4}+\rho_2-3/5\rho_3\\
 u_{4}  &  = & \rho_1\\
  5/k  &  \ge & -\rho_2-6/5\rho_3\\
\end{array}
\end{equation}
which solution, using the $\Gamma$-algorithm is:
\begin{equation}\label{e202}
\left(\begin{array}{c}
 F_{y_4} \cr  \rho_1 \cr  \rho_2 \cr  \rho_3 \cr
\end{array}\right)=\left(\begin{array}{c}
 551/807 \cr  0 \cr  -52441/(12912k) \cr  -60595/(77472k) \cr
\end{array}\right)+\pi\left(\begin{array}{c}
1 \cr  0 \cr  233/(80k) \cr  287/(48k) \cr
\end{array}\right);\;\;\pi\ge 0,
\end{equation}
which is composed of a particular solution and a cone generated by
a single vector. Note that $F_{y_4}=551/807$ is the force
necessary to get a vertical displacement of node $4$ equal to
$-5/k$. Since it cannot be smaller and must satisfy the constraint
$v_4\ge -5/k$, the value of $\pi$ has to be positive.

Replacing (\ref{e202}) in (\ref{equ2a}) we get:
\begin{equation}\label{equ27a}
\left(\begin{array}{c}
 u_{1} \cr  v_{1} \cr  u_{2} \cr  v_{2} \cr  u_{3} \cr  v_{3} \cr  u_{4} \cr  v_{4} \cr
\end{array}\right)=\fraca{1}{k}\left(\begin{array}{c}
 0 \cr  0 \cr  177/538 \cr  -81521/12912 \cr  1437/538 \cr  -52441/12912 \cr  0 \cr  -5 \cr
\end{array}\right)+\fraca{\pi}{k}\left(\begin{array}{c}
0 \cr  0 \cr  9/5 \cr  287/40 \cr  -9/5 \cr  233/80 \cr  0 \cr
807/80 \cr
\end{array}\right);\;\;\pi\ge 0,
\end{equation}
which, as before, is the sum of a particular solution and a cone
generated by a single vector.

 From an engineering point of view, this solution
must be interpreted as follows:
\begin{enumerate}
  \item The particular solution gives the displacements produced by the force $F_{y_4}=551/807$
   that is the minimum force required to satisfy the constraint $v_4\ge
-5/k$. Note that the vertical displacement of node $4$ is equal to
$-5/k$.

  \item The cone generator represents the
  displacement increments produced by a unit force acting on the lifting jack.
   Note that the vertical
  displacement at node $4$ is equal to $807/(80k)$, and that
  because the $\pi$-value has to be positive the vertical displacement of node
  $4$ is always greater than $-5/k$.

\item Note that both solutions satisfy the boundary conditions
$u_1=0$, $v_1=0$, $u_4=0$.
\end{enumerate}
}
\end{example}

\begin{example}[Maximum compression constraint]\label{ex29}{\em
Consider the same structure as in Figure {\ref{f15}}. The aim is to find all possible values of forces $P_1$, $P_2$, and $P_3$, in such a way that bars 1 and 5 are always subject to positive stresses (tractions). To solve this problem a
relation between the displacements $u_i^{(l)}, v_i^{(l)}$,
$u_j^{(l)},v_j^{(l)}$ and the axial force $N^{(l)}$ acting on bar
$l$ must be established. From (\ref{equ4}) and (\ref{equ8}) we
have:
\begin{equation}\label{equ28}
{\bf N}^{(l)}=k \left(
\begin{array}{cccc}
  -1 & 0 & 1 & 0 \\
\end{array}\right)
{\bf L}^{(l)}{\bf u}^{(l)} =k\left(\begin{array}{cccc}
  -\cos \alpha^{(l)} & -\sin \alpha^{(l)} & \cos \alpha^{(l)} & \sin \alpha^{(l)} \\
\end{array}\right)
\left(\begin{array}{c}
 u_i^{(l)}\\
 v_i^{(l)}\\
 u_j^{(l)}\\
 v_j^{(l)}\\
\end{array}\right).
\end{equation}

If bars $1$ and $5$ are to be subject to tensions ($N^{(1)}\ge 0$
and $N^{(5)}\ge 0$) and the structure must satisfy the boundary
conditions $u_1=0$, $v_1=0$ and $u_4=0$, the following system of
inequations obtained from (\ref{equ29}) and (\ref{equ28}) must be
solved:
\begin{equation}\label{equ30}
\begin{array}{rcl}
u_{1} &  = &  3P_1/(5k) +9P_2/(20k)+341P_3/(120k) +287P_4/(60k)+\rho_1+4/5\rho_3\\
v_{1} & = &  9P_1/(20k) +19P_2/(10k)+25P_3/(32k)-27P_4/(40k) +\rho_2-3/5\rho_3\\
u_{4}  &  = & \rho_1\\
0  &  \ge & 3P_3/4+3P_4/2\\
0  &  \ge & P_1 +3P_2/4+3P_3/2+3P_4/2,\\
\end{array}
\end{equation}
where the last two inequalities represent the constraints
$-N^{(1)}\le 0$ and $-N^{(5)}\le 0$, respectively.

The solution of the system using the $\Gamma$-algorithm is:
\begin{equation}\label{equ31}
\left(\begin{array}{c} P_1 \cr P_2 \cr P_3 \cr P_4 \cr  \rho_1 \cr
\rho_2 \cr \rho_3 \cr
\end{array}\right)=\renewcommand{\arraycolsep}{-0.03cm}\rho_4\left(\begin{array}{c}
 3/4 \cr  1 \cr  -2 \cr  1 \cr 0 \cr 0 \cr 0 \cr
\end{array}\right)+\rho_5\left(\renewcommand{\arraystretch}{1.8}\begin{array}{c}
 \fraca{12}{25} \cr  -\fraca{16}{25} \cr  0 \cr  0 \cr 0 \cr \fraca{1}{k} \cr 0 \cr
\end{array}\right)+\pi_1\left(\begin{array}{c}
 1 \cr  4/3 \cr  -4/3 \cr  0 \cr 0 \cr 0 \cr \fraca{233}{(72k)} \cr
\end{array}\right)+\pi_2\left(\renewcommand{\arraystretch}{1.8}\begin{array}{c}
 -\fraca{179}{125} \cr  \fraca{72}{125} \cr  0 \cr  0 \cr 0 \cr 0 \cr \fraca{3}{(4k)} \cr
\end{array}\right);\rho_4,\rho_5\in\R; \;\pi_1,\pi_2\ge 0
\end{equation}

Replacing (\ref{equ31}) in (\ref{equ29}) we get:
\begin{equation}\label{equ32}
\left(\begin{array}{c}
 u_{1} \cr  v_{1} \cr  u_{2} \cr  v_{2} \cr  u_{3} \cr  v_{3} \cr  u_{4} \cr  v_{4} \cr
\end{array}\right)=\fraca{\rho_4}{k}\left(\begin{array}{c}
 0 \cr  0 \cr  0 \cr  \fraca{25}{16} \cr  0 \cr  0 \cr  0 \cr  0 \cr
\end{array}\right)+\fraca{\rho_5}{k}\left(\begin{array}{c}
 0 \cr  0 \cr  0 \cr  1 \cr  0 \cr  1 \cr  0 \cr  1 \cr
\end{array}\right)+\fraca{\pi_1}{k}\left(\begin{array}{c}
0 \cr  0 \cr  1 \cr  \fraca{179}{60} \cr  0 \cr  0 \cr  0 \cr \fraca{233}{60} \cr
\end{array}\right)+\fraca{\pi_2}{k}\left(\begin{array}{c}
0 \cr  0 \cr  0 \cr  \fraca{9}{10} \cr  -\fraca{6}{5} \cr  0 \cr  0 \cr \fraca{9}{10} \cr
\end{array}\right);\rho_4,\rho_5\in\R; \pi_1,\pi_2\ge 0,
\end{equation}
which is a cone, i.e., the sum of a linear space of dimension
$2$ and an acute cone generated by two vectors.

\begin{figure}[h]
\begin{center}
\includegraphics*[width=\textwidth]{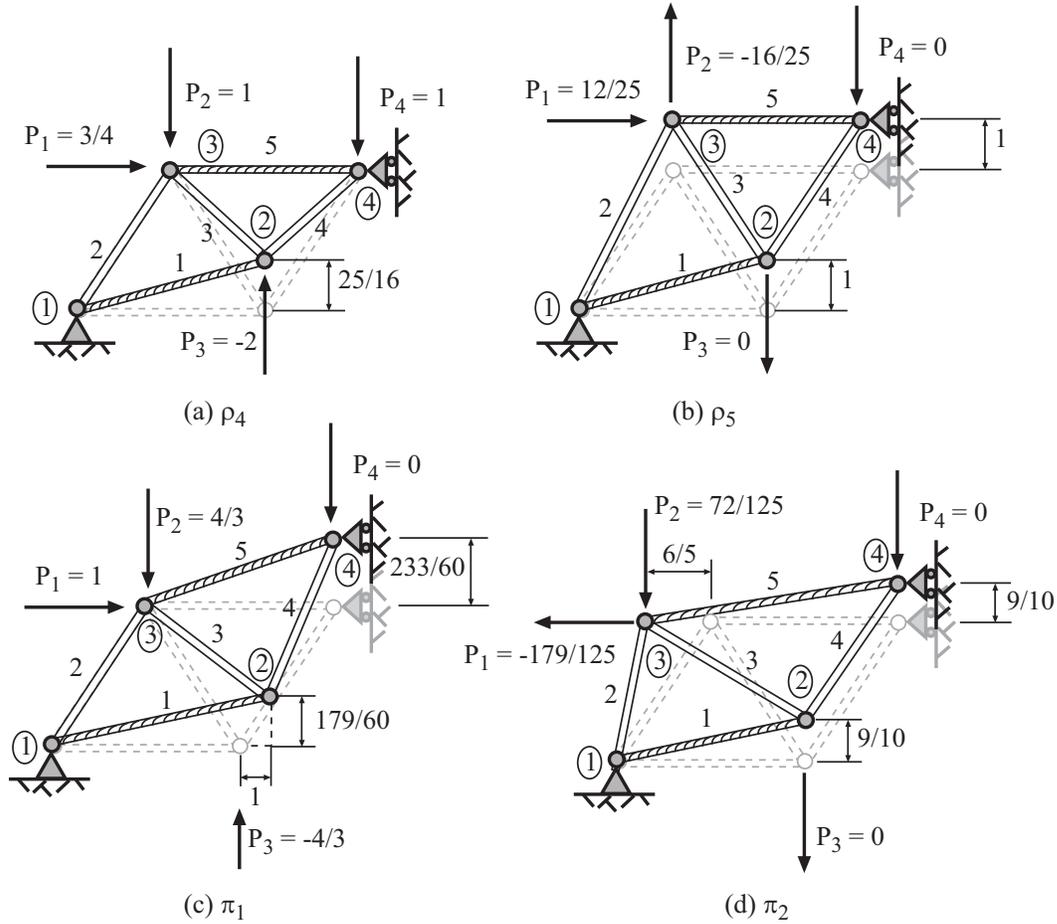}
\caption{Illustration of the different basic vectors that generate
the linear space and the cone components of the solution of
Example \ref{ex29}, where the tension bars 1 and 5 have been
distinguished from the remaining bars.}\label{f16}
\end{center}
\end{figure}

From an engineering point of view, this solution must be
interpreted as follows:
\begin{enumerate}
 \item The fist linear space generator in (\ref{equ32}) corresponds to a vertical displacement of
  node 2 with the remaining nodes being fixed, which
  leads to compressions in bars 3 and 4 and no stresses
  in the remaining bars (see Figure \ref{f16}(a)).

  \item  The second linear space generator in (\ref{equ32}) corresponds to a vertical displacement
   of nodes 2, 3 and 4 with node 1 remaining fixed, which
  corresponds to a rigid vertical displacement of the substructure defined by those nodes,
  while bar 2 is subject to tension stress.
  This implies that bars 1, 3, 4 and 5 are subject to no stress (see Figure \ref{f16}(b)).

  \item The first acute cone generator in (\ref{equ32}) corresponds to a rotation
with respect to node 3 of the substructure defined by nodes 2, 3
and 4, with the bar 2 remaining fixed, while bar 1 is subject to tension stress. This implies that bars 2, 3, 4 and 5 are subject to no stress (see Figure \ref{f16}(c)).

  \item The second acute cone generator in (\ref{equ32}) corresponds to a
  vertical translation of nodes 3 and 4
and a horizontal displacement of node 3. This implies that bar 2
is subject to compression, bar 5 to tension, and bars 1, 3 and 4
are subject to no stress (see Figure
\ref{f16}(d)).

\item Note that all vertices satisfy the boundary conditions
$u_1=v_1=u_4=0$.

\item Note also that the rotation, horizontal and vertical
displacements associated with the acute cone generators must have
the sign indicated in Figures \ref{f16}(c) and \ref{f16}(d)
($\pi_1,\pi_2\in\R^+)$ for the bars 1 and 5 being subject to positive stress (tension), while the vertical displacements
associated with the linear space generators can have any direction
($\rho_4,\rho_5\in\R$) because bars 1 and 5 do not work under these displacements.
\end{enumerate}

 }
\end{example}

\begin{example}[Tension, compression and deflection constraints]\label{ex30a}{\em
Consider the same structure as in Figure {\ref{f15}}, but now bars
$3$ and $4$ have been replaced by cables, that can only stand
tensions, and bars $1, 2$ and $5$ are built using concrete,
material specially suitable to support compressions.

The target of this example consists of obtaining the values of the
forces $P_1, P_2, P_3$ and $P_4$ that can be applied on the
structure in such a way that the cables work under tension and the
concrete beams under compression, respectively. In addition, the
maximum vertical deflection of node $4$ is limited to $v_2\ge
-1/k$.

If bars $3$, and $4$ are to be subject to tensions ($N^{(3)}\ge 0$
and $N^{(4)}\ge 0$), bars $1, 2$ and $5$ to compressions
($N^{(1)},N^{(2)}\le 0$ and $N^{(5)}\le 0$) and the structure must
satisfy the boundary conditions $u_1=0$, $v_1=0$ and $u_4=0$, the
following system of inequations obtained from (\ref{equ29}) and
(\ref{equ28}) must be solved:
\begin{equation}\label{equ30a}
\begin{array}{rcl}
u_{1} &  = &  3P_1/(60k) +9P_2/(20k)+341P_3/(120k) +287P_4/(60k)+\rho_1+4/5\rho_3\\
v_{1} & = &  9P_1/(20k) +19P_2/(10k)+25P_3/(32k)-27P_4/(40k) +\rho_2-3/5\rho_3\\
u_{4}  &  = & \rho_1\\
0  &  \ge & -3P_3/4-3P_4/4\\
0  &  \ge & -5P_2/4-5P_3/4-5P_4/4\\
0  &  \ge & -5P_3/4-5P_4/4\\
0  &  \ge & 5P_4/4\\
0  &  \ge & -P_1 -3P_2/4-3P_3/2-3P_4/2\\
1/k  &  \ge & 9P_1/20
+27P_2/80+233P_3/160+27P_4/40-\rho_2-3/5\rho_3,
\end{array}
\end{equation}
where the last six inequalities represent the constraints
$N^{(1)}\le 0$, $N^{(2)}\le 0$, $-N^{(3)}\le 0$, $-N^{(4)}\le 0$,
$N^{(5)}\le 0$ and $-v_2\le 1/k$, respectively.

The solution of the system using the $\Gamma$-algorithm is:
\begin{equation}\label{equ33}
\left(\begin{array}{c} P_1 \cr P_2 \cr P_3 \cr P_4 \cr \rho_1 \cr
\rho_2 \cr \rho_3 \cr
\end{array}\right)=\renewcommand{\arraycolsep}{0.00cm}\lambda_1\left(\renewcommand{\arraystretch}{1.8}\begin{array}{c}
 -\fraca{12}{25} \cr  \fraca{16}{25} \cr  \fraca{32}{25} \cr  -\fraca{16}{25} \cr 0 \cr 0 \cr 0 \cr
\end{array}\right)+\lambda_2\left(\renewcommand{\arraystretch}{1.8}\begin{array}{c}
 -\fraca{12}{15} \cr  \fraca{16}{25} \cr  0 \cr  0 \cr 0 \cr -1/k \cr 0 \cr
\end{array}\right)+\lambda_3\left(\renewcommand{\arraystretch}{1.8}\begin{array}{c}
 \fraca{5}{9} \cr  0 \cr  0 \cr  0 \cr 0 \cr -\fraca{1}{(2k)} \cr -\fraca{5}{(12k)} \cr
\end{array}\right)+\lambda_4\left(\renewcommand{\arraystretch}{1.8}\begin{array}{c}
 -\fraca{60}{179} \cr  -\fraca{80}{179} \cr  \fraca{80}{179} \cr  0 \cr 0 \cr 0 \cr -\fraca{1165}{(1074k)} \cr
\end{array}\right)
+\lambda_5\left(\begin{array}{c}
 0 \cr  0 \cr  0 \cr  0 \cr 0 \cr 0 \cr 0 \cr
\end{array}\right),
\end{equation}
where
\begin{equation}\label{equ35}
 \sum_{i=1}^5\lambda_i=1;\;\lambda_i\ge 0;\;i=1,\ldots,5,
\end{equation}
and replacing (\ref{equ33}) in (\ref{equ29}) we get:
\begin{equation}\label{equ34}
\left(\begin{array}{c}
 u_{1} \cr  v_{1} \cr  u_{2} \cr  v_{2} \cr  u_{3} \cr  v_{3} \cr  u_{4} \cr  v_{4} \cr
\end{array}\right)=\renewcommand{\arraycolsep}{0.00cm}\fraca{\lambda_1}{k}\left(\begin{array}{c}
0 \cr  0 \cr  0 \cr  -1 \cr  0 \cr  0 \cr  0 \cr 0 \cr
\end{array}\right)+\fraca{\lambda_2}{k}\left(\begin{array}{c}
0 \cr  0 \cr  0 \cr  -1 \cr  0 \cr  -1 \cr  0 \cr -1 \cr
\end{array}\right)+\fraca{\lambda_3}{k}\left(\begin{array}{c}
0 \cr  0 \cr  0 \cr  -1 \cr  2/3 \cr  -1/2 \cr  0 \cr -1 \cr
\end{array}\right)+\fraca{\lambda_4}{k}\left(\begin{array}{c}
0 \cr  0 \cr  -72/179 \cr  -1 \cr  0 \cr  0 \cr  0 \cr -233/179
\cr
\end{array}\right)+\fraca{\lambda_5}{k}\left(\begin{array}{c}
0 \cr  0 \cr  0 \cr  0 \cr  0 \cr  0 \cr  0 \cr 0 \cr
\end{array}\right),
\end{equation}
where $\sum_{i=1}^5\lambda_i=1;\;\lambda_i\ge 0;\;i=1,\ldots,5,$
which is the sum of a polytope generated by $5$ vectors (vertices).

\begin{figure}[h]
\begin{center}
\includegraphics*[width=\textwidth]{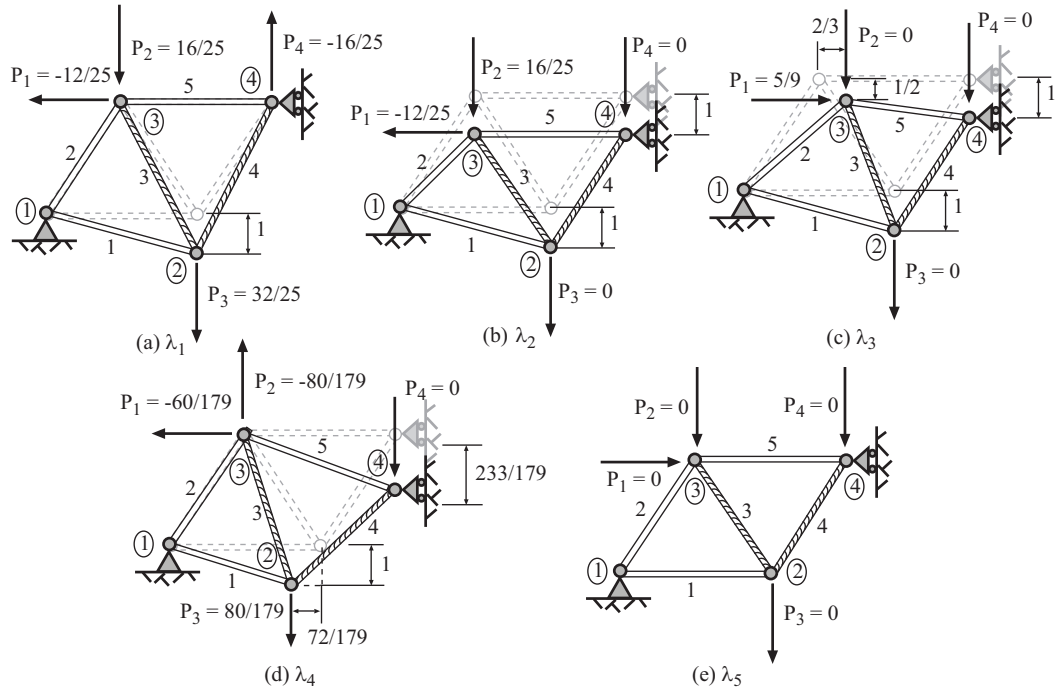}
\caption{Illustration of the different basic vectors that generate
the linear space and the cone components of the solution of
Example \ref{ex30a}.}\label{f17}
\end{center}
\end{figure}

From an engineering point of view, this solution must be
interpreted as follows:
\begin{enumerate}
  \item The fist vertex in (\ref{equ34}) corresponds to a vertical displacement of node 2 with the remaining nodes being fixed, which
  leads to traction stresses in bars 3 and 4 and no stress in the remaining bars (see Figure \ref{f17}(a)).

  \item  The second vertex in (\ref{equ34}) corresponds to a vertical displacement of nodes 2, 3 and 4 with node 1 remaining fixed, which
  corresponds to a rigid vertical displacement of the substructure defined by those nodes, while node 2 is subject to compression stress.
  This implies that bars 1, 3, 4 and 5 are to no stress (see Figure \ref{f17}(b)).

  \item The third vertex in (\ref{equ34}) corresponds to a vertical displacement of bar 4 and a rotation with respect to node 1 of the substructure
  defined by nodes 1, 2 and 3, which leads to a compression of bar
  5. This implies that bars 1, 2, 3 and 4 are subject to no stress (see Figure \ref{f17}(c)).

\item The fourth vertex in (\ref{equ34}) corresponds to a rotation
with respect to node 3 of the substructure defined by nodes 2, 3
and 4, with the bar 2 remaining fixed, while bar 1 is subject to compression stress. This implies that bars 2, 3, 4 and 5 are subject to no stress (see Figure \ref{f17}(d)).

\item The fifth vertex in (\ref{equ34}) corresponds to no
displacement at all, which implies no stress in
all bars  (see Figure \ref{f17}(e)).

\item Note that all vertices satisfy the boundary conditions
$u_1=v_1=u_4=0$.

\item Note also that the maximum allowed vertical displacement in node 2 ($v_2\ge -1/k$)
corresponds to all cases with $\lambda_5=0$, i.e., $v_2=-(\lambda_1+\lambda_2+\lambda_3+\lambda_4)=-1$.
\end{enumerate}
}
\end{example}

\section{Conclusions}\label{s8}The following conclusions can be
derived from this paper:
\begin{enumerate}
  \item A full understanding of real problems stated as systems of linear
  equations or inequalities requires both the mathematical and the engineering
  points of view that complement each other.

  \item The compatibility conditions must be interpreted from an
  engineering point of view, which help to identify errors,
  omissions or possible discrepancies between the mathematical
  model and the reality being modeled.

  \item The mathematical structures of the general solutions,
  linear spaces, cones, polytopes and mixed combinations of these
  three structures have clear engineering interpretations that
  are closely related to the real problem being modeled.

  \item The generators of the solution set, i.e., the linear space
  generators (basis), the cone generators, and the polytope
  generators (vertices) have clear interpretations from an engineering point
   of view, and contains a valuable information on the general
   solution of the problem and its properties.
\end{enumerate}

%\begin{center} {\large \bf Acknowledgments}\end{center}
%\noindent The authors are indebted to the Ministry of Science and Education
%of Spain through CICYT Projects BFM2003-05695 and BIA2005-07802-C02-01, to the Junta de Comunidades de Castilla-La
%Mancha through projects PAI-05-044 and PBI-05-053, and to the University of
%Castilla-La Mancha through project 0111001321 Program 541A for
%partial support of this work.

\newpage

\end{document}